\documentclass[useAMS,usenatbib,twocolumn]{mn2e}
\usepackage{amsmath}
\usepackage{amsfonts}
\usepackage{xcolor}
\usepackage[draft]{graphicx}

\makeatletter
\def\ref@jnl#1{{\rmfamily#1}}%
\newcommand\aj{\ref@jnl{AJ}}%
\newcommand\araa{\ref@jnl{ARA\&A}}%
\newcommand\apj{\ref@jnl{ApJ}}%
\newcommand\apjl{\ref@jnl{ApJ}}%
\newcommand\apjs{\ref@jnl{ApJS}}%
\newcommand\ao{\ref@jnl{Appl.~Opt.}}%
\newcommand\apss{\ref@jnl{Ap\&SS}}%
\newcommand\aap{\ref@jnl{A\&A}}%
\newcommand\aapr{\ref@jnl{A\&A~Rev.}}%
\newcommand\aaps{\ref@jnl{A\&AS}}%
\newcommand\azh{\ref@jnl{AZh}}%
\newcommand\baas{\ref@jnl{BAAS}}%
\newcommand\jcap{\ref@jnl{JCAP}}%
\newcommand\jrasc{\ref@jnl{JRASC}}%
\newcommand\memras{\ref@jnl{MmRAS}}%
\newcommand\mnras{\ref@jnl{MNRAS}}%
\newcommand\pra{\ref@jnl{Phys.~Rev.~A}}%
\newcommand\prb{\ref@jnl{Phys.~Rev.~B}}%
\newcommand\prc{\ref@jnl{Phys.~Rev.~C}}%
\newcommand\prd{\ref@jnl{Phys.~Rev.~D}}%
\newcommand\pre{\ref@jnl{Phys.~Rev.~E}}%
\newcommand\prl{\ref@jnl{Phys.~Rev.~Lett.}}%
\newcommand\pasp{\ref@jnl{PASP}}%
\newcommand\pasj{\ref@jnl{PASJ}}%
\newcommand\qjras{\ref@jnl{QJRAS}}%
\newcommand\skytel{\ref@jnl{S\&T}}%
\newcommand\solphys{\ref@jnl{Sol.~Phys.}}%
\newcommand\sovast{\ref@jnl{Soviet~Ast.}}%
\newcommand\ssr{\ref@jnl{Space~Sci.~Rev.}}%
\newcommand\zap{\ref@jnl{ZAp}}%
\newcommand\nat{\ref@jnl{Nature}}%
\newcommand\iaucirc{\ref@jnl{IAU~Circ.}}%
\newcommand\aplett{\ref@jnl{Astrophys.~Lett.}}%
\newcommand\apspr{\ref@jnl{Astrophys.~Space~Phys.~Res.}}%
\newcommand\bain{\ref@jnl{Bull.~Astron.~Inst.~Netherlands}}%
\newcommand\fcp{\ref@jnl{Fund.~Cosmic~Phys.}}%
\newcommand\gca{\ref@jnl{Geochim.~Cosmochim.~Acta}}%
\newcommand\grl{\ref@jnl{Geophys.~Res.~Lett.}}%
\newcommand\jcp{\ref@jnl{J.~Chem.~Phys.}}%
\newcommand\jgr{\ref@jnl{J.~Geophys.~Res.}}%
\newcommand\jqsrt{\ref@jnl{J.~Quant.~Spec.~Radiat.~Transf.}}%
\newcommand\memsai{\ref@jnl{Mem.~Soc.~Astron.~Italiana}}%
\newcommand\nphysa{\ref@jnl{Nucl.~Phys.~A}}%
\newcommand\physrep{\ref@jnl{Phys.~Rep.}}%
\newcommand\physscr{\ref@jnl{Phys.~Scr}}%
\newcommand\planss{\ref@jnl{Planet.~Space~Sci.}}%
\newcommand\procspie{\ref@jnl{Proc.~SPIE}}%

\title[A NILC analysis of WMAP $9$-year polarisation data]{A needlet ILC analysis
    of WMAP $9$-year polarisation data: CMB polarisation
    power spectra} \author[Soumen Basak, Jacques Delabrouille]
        {Soumen Basak$^{1,2}$\thanks{E-mail: basak@apc.univ-paris7.fr},
          Jacques Delabrouille$^{1}$\thanks{E-mail:
            delabrouille@apc.univ-paris7.fr}\\ $^1$  Laboratoire APC, CNRS \& Universit\'e Paris Diderot, 10 rue
          Alice Domon et L\'eonie Duquet, 75205 Paris Cedex 13,
          France\\
        $^2$ Laboratoire AIM, UMR CEA-CNRS-Paris 7, Irfu, SAp/SEDI, Service d'Astrophysique, CEA Saclay, F-91191 GIF-SUR-YVETTE\\ CEDEX, France}

\makeatletter

\begin{document}

\pagerange{\pageref{firstpage}--\pageref{lastpage}} \pubyear{2010}

\maketitle

\label{firstpage}


\begin{abstract}

  We estimate Cosmic Microwave Background (CMB) polarisation power
  spectra, and temperature-polarisation cross-spectra, from the
  $9$-year data of the Wilkinson Microwave Anisotropy Probe
  (WMAP). Foreground cleaning is implemented using minimum variance
  linear combinations of the coefficients of needlet decompositions of
  sky maps for all WMAP channels, to produce maps for CMB temperature
  anisotropies ($T$-mode) and polarisation ($E$-mode and $B$-mode),
  for $9$ different years of observation. The final power spectra are
  computed from averages of all possible cross-year power spectra
  obtained using foreground-cleaned maps for the different years.
  Our analysis technique yields a measurement of the $EE$ spectrum
  that is in excellent agreement with theoretical expectations from
  the current cosmological model. By comparison, the publicly
  available WMAP $EE$ power spectrum is higher on average (and
  significantly higher than the predicted $EE$ spectrum from the
  current best fit) at scales larger than about a degree, an excess
  that is not confirmed by our analysis. Our $TE$ and $TB$
  measurements are in good agreement overall with the WMAP ones and
  are compatible with the theoretical expectations, although a few
  data points are off by a few standard deviations, and yield a
  reduced $\chi^2$ somewhat above expectation. As predicted for a
  standard cosmological model with low tensor to scalar ratio, the
  $EB$ and $BB$ power spectra obtained in our analysis are compatible
  with zero.

\end{abstract}



\section{Introduction}
The Cosmic Microwave Background (CMB), relic radiation emitted when
our Universe was about 380,000 years old, provides direct information
about the origin and history of cosmic structure. Hence, the current
understanding of cosmological evolution is heavily based on
observations of the CMB. Over the last two decades, many experiments
have accumulated observations of its temperature anisotropies,
imprints of the original density perturbations that later on gave rise
to the large scale structures observable today. Two very successful
space missions, the Cosmic Background Explorer \citep[COBE,
][]{1996ApJ...464L...1B}, and the Wilkinson Microwave Anisotropy Probe
\citep[WMAP, ][]{2003ApJS..148....1B}, complemented by several
ground-based and balloon-borne experiments, such as CAT
\citep{1999MNRAS.308.1173B}, {\sc Boomerang}
\citep{2000Natur.404..955D}, {\sc Maxima} \citep{2000ApJ...545L...5H},
{\sc Archeops} \citep{2003A&A...399L..19B,2003A&A...399L..25B}, CBI
\citep{2003ApJ...591..556P}, VSA
\citep{2004MNRAS.353..732D,2004MNRAS.353..747R}, ACBAR
\citep{2009ApJ...694.1200R}, ACT
\citep{2011ApJ...729...62D,2011PhRvL.107b1301D}, and SPT
\citep{2011ApJ...743...28K}, to mention just a few, have measured the
CMB temperature anisotropies at various angular scales and at various
wavelengths. Recently, the Planck
collaboration has released a precise measurement of the CMB
temperature power spectrum \citep{2013arXiv1303.5075P} which led to an
update of the best fit cosmological parameters
\citep{2013arXiv1303.5076P}.

The CMB angular power spectra measured by these experiments strongly
support a present cosmological model in which the Universe is
spatially flat, with an energy density dominated by the presence of
about 70\% dark energy and about 30\% matter, and in which structure
forms from gravitational collapse of primordial adiabatic
perturbations in the density of the cosmological fluid. At present,
this model does not require the additional admixture of primordial
tensor perturbations (gravitational waves), although cosmological
scenarios generically predict their existence. The observed CMB
anisotropies are connected to the state of the cosmological fluid at
recombination in a way that depends on a small set of parameters
specific of the cosmological model. CMB observations are of prime
importance for constraining these parameters of the model, as well as
for testing its internal consistency.

Temperature anisotropies alone, however, do not provide the complete
picture of the Universe. Independent information is needed to lift
degeneracies between cosmological parameter sets compatible with the
CMB temperature power spectrum. Other cosmological probes such as the
direct measurement of the present expansion rate
\citep{2001ApJ...553...47F}, the observation of the expansion history
with supernova
\citep{1999ApJ...517..565P,2006A&A...447...31A,2010A&A...523A...7G},
or baryonic oscillations traced by the distribution of galaxies
\citep{2005ApJ...633..560E}, provide complementary constraints
\citep{2006JCAP...10..014S}.

On the side of the CMB itself, additional informations are obtained by
measuring CMB polarisation. The CMB is indeed partially polarised due
to Thomson scattering of quadrupolar distribution of photons on free
electrons at the time of recombination \citep{1968ApJ...153L...1R}, at
a redshift $z_{\rm rec}$ of about $1100$. Large scale polarisation
also arises from the scattering of CMB photons on free electrons after
reionisation of the Universe at much lower redshift ($z_{\rm
  reion}<10$). CMB polarisation is an additional observable, which
helps disentangling Sachs-Wolfe from Doppler contributions to the
anisotropies, makes possible the estimation of individual
contributions to the power spectra from scalar and tensor
perturbations, and also helps constraining the epoch of reionisation.
The amplitude of the polarisation, however, is significantly lower
than that of temperature anisotropies, which makes its precise
characterisation challenging.

CMB polarisation was first detected at sub-degree angular scales by
the DASI ground-based interferometer \citep{2005ApJ...624...10L}. It
was subsequently measured by {\sc Boomerang}
\citep{2006ApJ...647..813M,2006ApJ...647..833P}, {\sc Maxipol}
\citep{2007ApJ...665...55W}, CBI \citep{2007ApJ...660..976S}, CAPMAP
\citep{2008ApJ...684..771B}, QUaD
\citep{2009ApJ...692.1247P,2009ApJ...705..978B}, BICEP
\citep{2010ApJ...711.1123C}, QUIET \citep{2011ApJ...741..111Q}, and
the Wilkinson Microwave Anisotropy Probe \citep[WMAP,
][]{2009ApJS..180..296N,2011ApJS..192...16L}.

The level of the CMB polarisation (at most a few per cent, depending
on the angular scale), however, makes it easily contaminated by
foreground emissions from both the Galaxy and extragalactic
sources. Galactic emission, in particular, is polarised at the level
of a few tens of per cent, i.e. ten times more than CMB
anisotropies. The signal to foreground ratio is thus less favourable
for polarisation than for intensity, in particular on large angular
scales where Galactic synchrotron and dust emissions are
strong. Contamination by foregrounds would result in CMB polarisation
power spectra larger than predicted by the current cosmological
model. For WMAP, the main contaminant is synchrotron emission, but
even dust, which has been measured to be polarised at a level that can
exceed 10\% \citep{2004A&A...424..571B}, is a potential worry.

Hence, contamination from polarised foreground signals has to be
removed as much as possible for the accurate measurement of the
temperature and polarisation angular power spectrum of the CMB. The
effectiveness of a foreground cleaning technique is typically
significantly improved by the use of prior knowledge of the emission
properties of the contaminant (such as the frequency dependence,
typical angular power spectrum of its emission, or availability of
external templates). However, since the properties of foreground
contamination are poorly known for polarisation (and, in particular,
no good template of polarised foregrounds is presently available), it
is particularly crucial to develop and use data analysis tools that
use only the minimum possible prior assumptions about foreground
polarisation.

Among the possible methods for CMB cleaning, the so-called Internal
Linear Combination (ILC) method, first proposed for foreground
cleaning in the analysis of COBE-DMR data \citep{1992ApJ...396L...7B},
and discussed subsequently by many authors \citep[see, e.g.,
][]{1998ApJ...502....1T}, is a simple and effective way to combine
multifrequency observations to extract the CMB while rejecting
contamination by superimposed foreground signals.  This method is
based on two reasonably safe assumptions. The first is that the
amplitude of CMB emission is frequency independent in thermodynamic
units, i.e. that the CMB emission law scales in frequency as the
derivative of the blackbody spectrum of the cosmic background. The
second is that CMB fluctuations are not correlated to foreground
signals. Under these assumptions, the ILC estimates the CMB as a
linear combination of sky maps such that the variance of the estimate
is minimum, while preserving unit response to the CMB (see, however,
the appendix of \citet{2009A&A...493..835D} for a discussion of second
order corrections and biases, and \citet{2010MNRAS.401.1602D} for a
discussion of the impact of calibration errors).

Component separation with an ILC method can be straightforwardly
implemented either in real space or in harmonic space
\citep{2003PhRvD..68l3523T,2004ApJ...612..633E,2006ApJ...645L..89S,2006NewAR..50..854S,2008PhRvD..78b3003S,2011ApJ...739L..56S,2011BASI...39..163S}.
Here, as in our previous work
\citep{2009A&A...493..835D,2012MNRAS.419.1163B}, we instead implement
the ILC on a frame of spherical wavelets, called needlets. This
special type of wavelets on the sphere provides good localisation in
both pixel space and harmonic space because they have compact support
in the harmonic domain, while still being very well localised in the
pixel domain \citep{narcowich:petrushev:ward:2006,2008MNRAS.383..539M,
  guilloux2009}. Needlets have already been used in various analyses
of WMAP data besides component separation and power spectrum
estimation, for instance by \citet{2008PhRvD..78j3504P} to detect
features in the CMB, and by \cite{2009ApJ...701..369R} to put limits
on the non-Gaussianity parameter $f_{\rm NL}$.

Recently, we have used a needlet ILC on WMAP $7$-year data to obtain
an estimate of the CMB temperature angular power spectrum
\citep{2012MNRAS.419.1163B}. In the present paper, as a natural
extension to this work, we address the problem of measuring the CMB
polarisation power spectra, and temperature-polarisation cross-spectra
from WMAP $9$-year observations as precisely as possible.

The paper is organised as follows: In section~\ref{sec:needlet-ilc} we
describe the methodology to estimate the CMB using an ILC on wavelet
decompositions of multi-frequency sky maps. The implementation of this
on WMAP $9$-year data, and the results for polarisation power spectra,
are described in sections \ref{sec:wmap-map}, \ref{sec:wmap-spec} and
\ref{sec:gof}. We conclude in section~\ref{sec:conclusion}.

\section{Needlet ILC estimate of CMB}
\label{sec:needlet-ilc}

A scalar field on the sphere such as CMB temperature anisotropies is
conveniently expanded in usual spherical harmonics:
\begin{eqnarray}
T(\hat n)=\sum _{l=0}^{\infty}\sum_{m=-l}^l T_{lm}Y_{l m}(\hat n)
\end{eqnarray}
The CMB polarisation field is usually specified using the Stokes
parameters, $Q$ and $U$, with respect to a particular choice of a
coordinate system on the sky in relation to which the linear
polarisation is defined. One can conveniently combine the Stokes
parameters into the single complex quantity, $P_\pm =Q\pm iU$. Due to
its rotation properties, one may expand $P_\pm(\hat n)$ in terms of
spin-$2$ spherical harmonics, ${}_{\pm 2}Y_{lm}(\hat n)$
\citep{1967JMP.....8.2155G}, as:
\begin{eqnarray}
P_\pm (\hat n)=\sum _{l=0}^{\infty}\sum_{m=-l}^l P_{\pm 2,lm}\,{}_{\pm 2}Y_{l m}(\hat n).
\end{eqnarray}
The description of CMB polarisation, however, traditionally makes use
of two scalar fields that are independent of how the coordinate system
is oriented, and are related to the Stokes parameters by a non-local
transformation \citep{1997PhRvD..55.1830Z,1997PhRvD..55.7368K}. One
of the fields, traditionally denoted as $E$, has even parity, whereas
the other one, $B$, has odd parity (and hence is pseudo-scalar, rather
than scalar). In harmonic space, the $E$ and $B$ modes of CMB
polarisation are related to the complex polarisation fields $P_\pm$
as,
\begin{eqnarray}
 P_{+2,lm}=-(E_{lm} + i B_{lm})
\end{eqnarray}
and
\begin{eqnarray}
 P_{-2,lm}=-(E_{lm} - i B_{lm}).
\end{eqnarray}

Hence, one can fully characterise CMB anisotropies using three
Gaussianly distributed random scalar fields ($T$, $E$ and $B$),
without loss of information. In the following, maps of $Q$ and $U$ are 
converted into maps of $E$ and $B$ by expansion onto
spin-$2$ spherical harmonics, followed by an inverse spherical
harmonic transform for $E_{lm}$ and $B_{lm}$ independently.

\subsection{The CMB data model}
Denote $X^{\text{OBS},c}(\hat{n})$, $(X={T,E,B})$ full-sky,
multi-frequency temperature anisotropy and polarisation maps of the
sky in $n_{c}$ different frequency bands (channels), such as those
provided by WMAP. The observed signal $X^{\text{OBS},c}(\hat{n})$ in
channel $c$ can be modelled as,
\begin{eqnarray}
X^{\text{OBS},c}(\hat{n})= \int_{\hat{n}^{\prime}} d\Omega_{\hat{n}^{\prime}}
\,\,b^c(\hat{n}.\hat{n}^{\prime})\,\,X^{\text{SIG},c}(\hat{n}^\prime) 
+ X^{\text{N},c}(\hat{n})
\label{eq:map_pix_obs}
\end{eqnarray}
where $X^{\text{SIG},c}(\hat{n})$ is the signal (sky) component,
itself decomposed in the sum of CMB and foreground components,
\begin{eqnarray}
X^{\text{SIG},c}(\hat{n})=a^{c}\,X^{\text{CMB}}(\hat{n})+X^{\text{FG},c}(\hat{n}),
\end{eqnarray}
$a^{c}$ being the CMB calibration coefficient for the channel $c$. Up
to calibration uncertainties, $a^c=1$ for all WMAP channels. If, in
addition to WMAP data, we use external data sets which serve as
foreground templates to help foreground subtraction, as done in the
present work, the coefficients $a^{c}$ vanish for such data sets
(i.e. the ancillary maps contain no CMB anisotropies).

The beam function $b^{c}(\hat{n}.\hat{n}^{\prime})$, represents the
smoothing of the signal due to the finite resolution of the
observations. Assuming for simplicity that the beams are circularly
symmetric (a good approximation for WMAP data),
$b^{c}(\hat{n}.\hat{n}^{\prime})$ depends only on the angle
$\theta=\cos^{-1}(\hat{n}.\hat{n}^{\prime})$ between the directions
$\hat{n}$ and $\hat{n}^{\prime}$, and can be expanded in terms of
Legendre polynomials,
\begin{eqnarray}
b^c(\hat{n}.\hat{n}^{\prime})=
\sum_{l=0}^{\infty}\frac{2l+1}{4\pi}b_{l}^cP_{l}(\hat{n}.\hat{n}^{\prime}).
\end{eqnarray}

The last term, $X^{\text{N},c}(\hat{n})$, in equation
(\ref{eq:map_pix_obs}) represents the detector noise in channel $c$, and
is not affected by the beam function.
For a spherically symmetric beam, equation \ref{eq:map_pix_obs} can be
recast straightforwardly in the spherical harmonic representation, as:
\begin{eqnarray}
X_{l m}^{\text{OBS},c}=a^{c}\,b_l^c\,X_{l m}^{\text{CMB}}+b_l^c\,X_{l
  m}^{\text{FG},c}+X_{l m}^{\text{N},c}
\label{map_harm_obs}
\end{eqnarray}
where $X_{lm}$ stands for the three modes $T_{lm}$, $E_{lm}$ and
$B_{lm}$ of temperature and polarisation in harmonic space.

\subsection{Implementation of the needlet transform}
Considering that each channel observes the sky at a different
resolution, the maps are first convolved/deconvolved, in harmonic
space, to the same resolution:
\begin{eqnarray}
X_{l m}^{c}=\frac{b_l}{b_l^c}\,\,X_{l m}^{\text{OBS},c}.
\label{map_harm}
\end{eqnarray}
Each of these maps $X_{l m}^{c}$ is then decomposed into a set of
filtered maps $X_{l m}^{c,j}$ represented by the spherical harmonic
coefficients,
\begin{eqnarray}
X_{l m}^{c,j}=h_{l}^{j}X_{l m}^{c},
\end{eqnarray}
where the filters $h_{l}^{j}$, serving for localisation in the
harmonic space, are chosen in such a way that
\begin{eqnarray}
\sum_{j}\left(h_{l}^{j}\right)^{2}=1.
\end{eqnarray}
The reconstruction of the original maps $X_{lm}^{c}$ from the
collection of the filtered maps $X_{l m}^{c,j}$, representing each a
different scale, is performed using the same set of filters.
In terms of $h_{l}^{j}$, the spherical needlets are defined as,
\begin{eqnarray}
\Psi_{j k}(\hat n)=\sqrt{\lambda_{jk}}\sum _{l=0}^{l_{\max }}\sum_{m=-l}^l 
h_{l}^{j}\,Y_{l m}^{*}(\hat n)\,Y_{l m}(\hat\xi_{jk}),
\end{eqnarray}
where $\{\xi_{jk}\}$ denote a set of cubature points on the sphere for
scale $j$. In practice, we identify these points with the pixel
centres in the HEALPix\footnote{http://healpix.jpl.nasa.gov}
pixelisation scheme \citep{2005ApJ...622..759G}. Each index $k$
corresponds to a particular HEALPix pixel, at a resolution parameter
{\tt nside}$(j)$ specific to that scale $j$. The cubature weights
$\lambda_{jk}$ are inversely proportional to the number $N_{j}$ of
pixels used for the needlet decomposition,
i.e. $\lambda_{jk}=\frac{4\pi}{N_{j}}$.
The needlet coefficients for CMB fields $X(\hat n)$ are denoted as,
\begin{eqnarray}
\beta^{X}_{j k}&=&\int_{S^{2}}X(\hat n)\,\Psi_{j k}(\hat
n)\,d\Omega_{\hat n} \nonumber\\ &=&\sqrt{\lambda_{j k}} \sum
_{l=0}^{l_{\max }} \sum_{m=-l}^l h_l^j\,b_l\,X_{l m}\,\,Y_{l m}(\xi
_{j k}).
\end{eqnarray}
The linearity of the needlet decomposition implies that the needlet
coefficients $\beta_{j k}^{c}$ corresponding to the filtered map
obtained from the harmonic coefficients $X_{l m}^{c,j}$ are a linear
combination of the needlet coefficients of individual components and
noise at HEALPix grid points $\xi_{j k}$:
\begin{eqnarray}
\beta_{j k}^{X,c}=a^{c}\,\beta_{j k}^{\text{CMB}}+\beta_{jk}^{\text{FG},c}+
\beta_{j k}^{\text{N},c}
\end{eqnarray}
where,
\begin{eqnarray}
\beta_{j k}^{\text{CMB}}&=&\sqrt{\lambda_{j k}} \sum _{l=0}^{l_{\max }}
\sum_{m=-l}^l h_l^j\,b_l\,X_{l m}^{\text{CMB}}\,\,Y_{l m}(\xi _{j
  k})\nonumber\\
\beta_{j k}^{\text{FG},c}&=&\sqrt{\lambda_{j k}} \sum _{l=0}^{l_{\max }}
\sum_{m=-l}^l h_l^j\,b_l\,X_{l m}^{\text{FG},c}\,\,Y_{l m}(\xi _{j
  k})\nonumber\\
\beta_{j k}^{\text{N},c}&=&\sqrt{\lambda_{j k}} \sum _{l=0}^{l_{\max }}
\sum_{m=-l}^l h_l^j\,\frac{b_l}{b_l^c}\,X_{l m}^{\text{N},c}\,\,Y_{l m}(\xi _{j
  k})
\end{eqnarray}

\subsection{Implementation of the needlet ILC}
The ILC estimate of needlet coefficients of the cleaned map is
obtained as a linearly weighted sum of the needlet coefficients
$\beta_{j k}^{c}$,
\begin{eqnarray}
\beta_{j k}^{\text{NILC}}=\sum_{c=1}^{n_c}\omega_{j k}^{c}\,\beta_{j k}^{X,c}
\end{eqnarray}
where $\omega_{j k}^{c}$ is the needlet weight for scale $j$ and
frequency channel $c$, at the pixel $k$ of the HEALPix representation
of the needlet coefficients for that scale.
Under the assumption of decorrelation between CMB and foregrounds, and
between CMB and noise, the empirical variance of the error is minimum
when the empirical variance of the ILC map itself is minimum.
The condition for preserving the CMB signal during the cleaning is
encoded as the constraint:
\begin{eqnarray}
\sum_{c=1}^{n_{c}}a^{c}\omega_{j k}^{c}=1.
\label{constraints}
\end{eqnarray}
The resulting needlet ILC weights $\widehat{\omega}_{j k}^{c}$ that
minimise the variance of the reconstructed CMB, subject to the
constraint that the CMB is preserved, are expressed as:
\begin{eqnarray}
\widehat{\omega}_{j
  k}^{c}=\frac{\sum_{c^{\prime}} \left[\widehat{R}_{j
    k}^{-1}\right]^{c
    c^{\prime}}a^{c^{\prime}}}{\sum_c \sum_{c^{\prime}}a^{c}
  \left[\widehat{R}_{j k}^{-1}\right]^{c c^{\prime}}a^{c^{\prime}}},
    \label{eq:ilc}
\end{eqnarray}
where indices $c$ and $c'$ of elements of the matrix
$\left[\widehat{R}_{j k}^{-1}\right]$ and of the vectors
$\widehat{\omega}_{jk}$ and $a$ are written down explicitly for
clarity. More compactly, we have:
\begin{eqnarray}
\widehat{\omega}_{j
  k}=\frac{\left[\widehat{R}_{j
    k}^{-1}\right]a}{a^{T}
  \left[\widehat{R}_{j k}^{-1}\right]a},
    \label{eq:ilc-matrix-form}
\end{eqnarray}
where $\widehat{\omega}_{jk}$ is the vector of ILC weights to be
applied to the needled coefficients of all input observations at scale
$j$ and in pixel $k$, $a$ is the CMB `mixing vector' (a vector of
$n_c$ entries all equal to unity for inputs in thermodynamic
temperature), and $\left[\widehat{R}_{jk}^{-1}\right]$ an estimate of
the inverse covariance of the needlet coefficients of the $n_c$
observations at pixel $k$ of scale $j$.

The NILC estimate of the cleaned CMB needlet coefficients is:
\begin{eqnarray}
\beta_{jk}^{\text{NILC}} = \beta_{jk}^{\text{CMB}}+ \sum_{c}
\widehat{\omega}_{jk}^{c} \left(\beta_{j
  k}^{\text{FG},c}+\beta_{jk}^{\text{N},c}\right).
    \end{eqnarray}

The elements of the covariance matrix for scale $j$ at pixel $k$,
$R_{j k}^{c c^{\prime}}=\left<\beta_{j k}^{c}\beta_{j
  k}^{c^{\prime}}\right>$, are obtained each as an average of the
product of the relevant computed needlet coefficients over some space
domain ${\cal D}_{k}$ centred at $k$. In practice, they are computed
as
\begin{eqnarray}
\widehat{R}_{X,j k}^{c c^{\prime}}=\frac{1}{n_{k}}\sum_{k^{\prime}}
w_j(k,k^\prime)\,\beta_{j k}^{X,c}\beta_{j k}^{X,c^{\prime}},
\end{eqnarray}
where the weights $w_j(k,k^\prime)$ define the domain ${\cal D}_{k}$.
A sensible choice is for instance $w_j(k,k^\prime) = 1$ for $k^\prime$
closer to $k$ than some limit angle, and $w_j(k,k^\prime) = 0$
elsewhere, or alternatively, $w_j(k,k^\prime)$ shaped as a Gaussian
beam of some given size that depends on the scale $j$ (which is what
we do here).

Finally, the NILC estimate of the cleaned CMB map can be reconstructed
from cleaned CMB needlet coefficients using the same set of filters
that was used to decompose the original maps into their needlet
coefficients. The NILC CMB  map is then
\begin{eqnarray}
X^{\text{NILC}}(\hat{n})&=&\sum_{l m} X_{l m}^{\text{NILC}}\,Y_{l
  m}(\hat{n})
\label{equ:cmb-nilc}
\end{eqnarray}
with
\begin{eqnarray}
X_{l m}^{\text{NILC}} = b_{l}\,X_{l
  m}^{\text{CMB}}+X_{l m}^{\text{RFG}}+X_{l
  m}^{\text{RN}},
\label{equ:cmb-nilc-cont}
\end{eqnarray}
where the harmonic coefficients residual foreground ($X_{l
  m}^{\text{RFG}}$) and residual noise ($X_{l m}^{\text{RN}}$) are
given by:
\begin{eqnarray}
X_{l m}^{\text{RFG}}=\sum_{j}\sum_{k}\sqrt{\lambda_{j k}}\,\beta_{j
  k}^{\text{RFG}}\,h_{l}^{j}\,Y_{l m}(\xi _{j k})
\end{eqnarray}
and
\begin{eqnarray}
X_{l m}^{\text{RN}}=\sum_{j}\sum_{k}\sqrt{\lambda_{j k}}\,\beta_{j
  k}^{\text{RN}}\,h_{l}^{j}\,Y_{l m}(\xi _{j k}).
\end{eqnarray}
Equations \ref{equ:cmb-nilc} and \ref{equ:cmb-nilc-cont} imply that
the NILC estimate of CMB contains some residual foreground and noise
contamination.

\section{WMAP $9$-year needlet ILC map}
\label{sec:wmap-map}
The WMAP satellite has observed the sky in five frequency bands
denoted K, Ka, Q, V and W, centred at $23$, $33$, $41$, $61$ and $94$ GHz
respectively. After $9$ years of observation, the released data
includes temperature anisotropy and polarisation maps obtained with
ten difference assemblies, for $9$ individual years. One map is
available, per year, for each of the K and Ka bands, two for the Q
band, two for the V band and four for the W band. These sky maps are
sampled using the HEALPix pixelisation scheme at a resolution level
(nside$=512$), corresponding to approximately $3$ million sky pixels.

We work on band-averaged maps of $T$, $E$ and $B$ for the five
frequency bands, complemented, for temperature only, by three
foreground templates (dust at $100$ microns, as obtained by
\citet{1998ApJ...500..525S}, the $408$ MHz synchrotron map of
\citet{1981A&A...100..209H}, and the composite all-sky H-alpha map of
\citet{2003ApJS..146..407F}). All sky maps are convolved/deconvolved
in harmonic space, to a common beam resolution (full width at half
maximum (FWHM)$=13.2$).

\begin{figure}
  \centering
  \includegraphics[scale=0.27,angle=90]{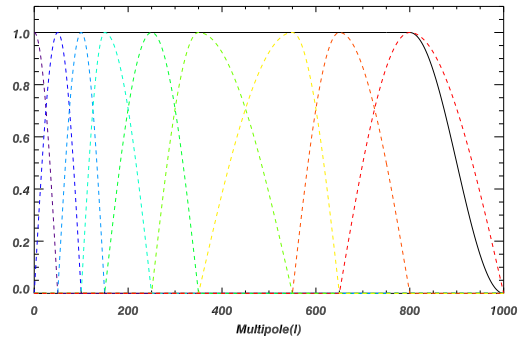}
  \caption{The broken lines show the needlet bands used in the present
    analysis. The solid black line shows the normalisation of the
    needlet bands, i.e. the total filter applied to the original map
    after needlet decomposition and synthesis of the output map from
    needlet coefficients.}
  \label{fig:needlet-bands}
\end{figure}

Each of these maps is then decomposed into a set of needlet
coefficients. For each scale $j$, needlet coefficients of a given map
are stored in the format of a single HEALPix map at degraded
resolution. The filters $h^{j}_{l}$ used to compute filtered maps are
shaped as follows:
\begin{eqnarray*} 
h^{j}_{l} = \left\{
\begin{array}{rl} 
\cos\left[\left(\frac{l^{j}_{peak}-l}{l^{j}_{peak}-l^{j}_{min}}\right)
\frac{\pi}{2}\right]& \text{for } l^{j}_{min} \le l < l^{j}_{peak},\\ 
\\
1\hspace{0.5in} & \text{for } l = l_{peak},\\
\\
\cos\left[\left(\frac{l-l^{j}_{peak}}{l^{j}_{max}-l^{j}_{peak}}\right)
\frac{\pi}{2}\right]& \text{for } l^{j}_{peak} < l \le l^{j}_{max} 
\end{array} \right. 
\end{eqnarray*}
For each scale $j$, the filter has compact support between the
multipoles $l^{j}_{min}$ and $l^{j}_{max}$ with a peak at
$l^{j}_{peak}$ (see figure \ref{fig:needlet-bands} and table
\ref{tab:needlet-bands}). The needlet coefficients $\beta^{X}_{j k}$
are computed from these filtered maps on HEALPix grid points $\xi_{j
  k}$ with resolution parameter nside equal to the smallest power of
$2$ larger than $l^{j}_{max}/2$.
\begin{table}
\caption{List of needlet bands used in the present
  analysis.}  \centering \begin{tabular}{c c c c c} \hline\hline
  Band index & $l_{min}$ & $l_{peak}$ & $l_{max}$ & nside \\ [5ex]
  \hline 1 & 0 & 0 & 50 & 32\\ 2 & 0 & 50 & 100 & 64 \\ 3 & 50 & 100 &
  150 & 128 \\ 4 & 100 & 150 & 250 & 128 \\ 5 & 150 & 250 & 350 & 256
  \\ 6 & 250 & 350 & 550 & 512 \\ 7 & 350 & 550 & 650 & 512 \\ 8 & 550
  & 650 & 800 & 512 \\ 9 & 650 & 800 & 1000 & 512 \\ [1ex] \hline
\end{tabular} 
\label{tab:needlet-bands} 
\end{table}
The estimates of needlet coefficients covariance matrices, for each
scale $j$, are computed by smoothing maps of products of needlet
coefficient $\beta^{c}_{j k}\beta^{c^{\prime}}_{j k}$ with Gaussian
beams. In this way, an estimate of needlet covariances at each point
$k$ is obtained as a local, weighted average of nearby needlet
coefficient products. The full width at half maximum (FWHM) of each of
the Gaussian windows used for this purpose is chosen to ensure the
computation of the statistics by averaging about $1200$ samples or more,
resulting from a trade-off between the localisation of the estimates
(which requires small windows), and the accuracy of the estimate
(which require large windows). Choosing a smaller FWHM results in
inaccuracy in the covariance estimates, and hence ILC bias. Choosing a
larger FWHM results in less localisation, and hence loss of
effectiveness of the needlet approach.

Using these covariance matrices, ILC weights are computed for each of
$T$, $E$ and $B$, for each scale $j$ and for each pixel $k$ of the
needlet representation at scale $j$. For each of $T$, $E$ and $B$, a
full sky CMB map, at the resolution of the WMAP W channel, is
synthesised from the NILC needlet coefficients.
\begin{figure}
  \centering
  \includegraphics[scale=0.3,angle=90]{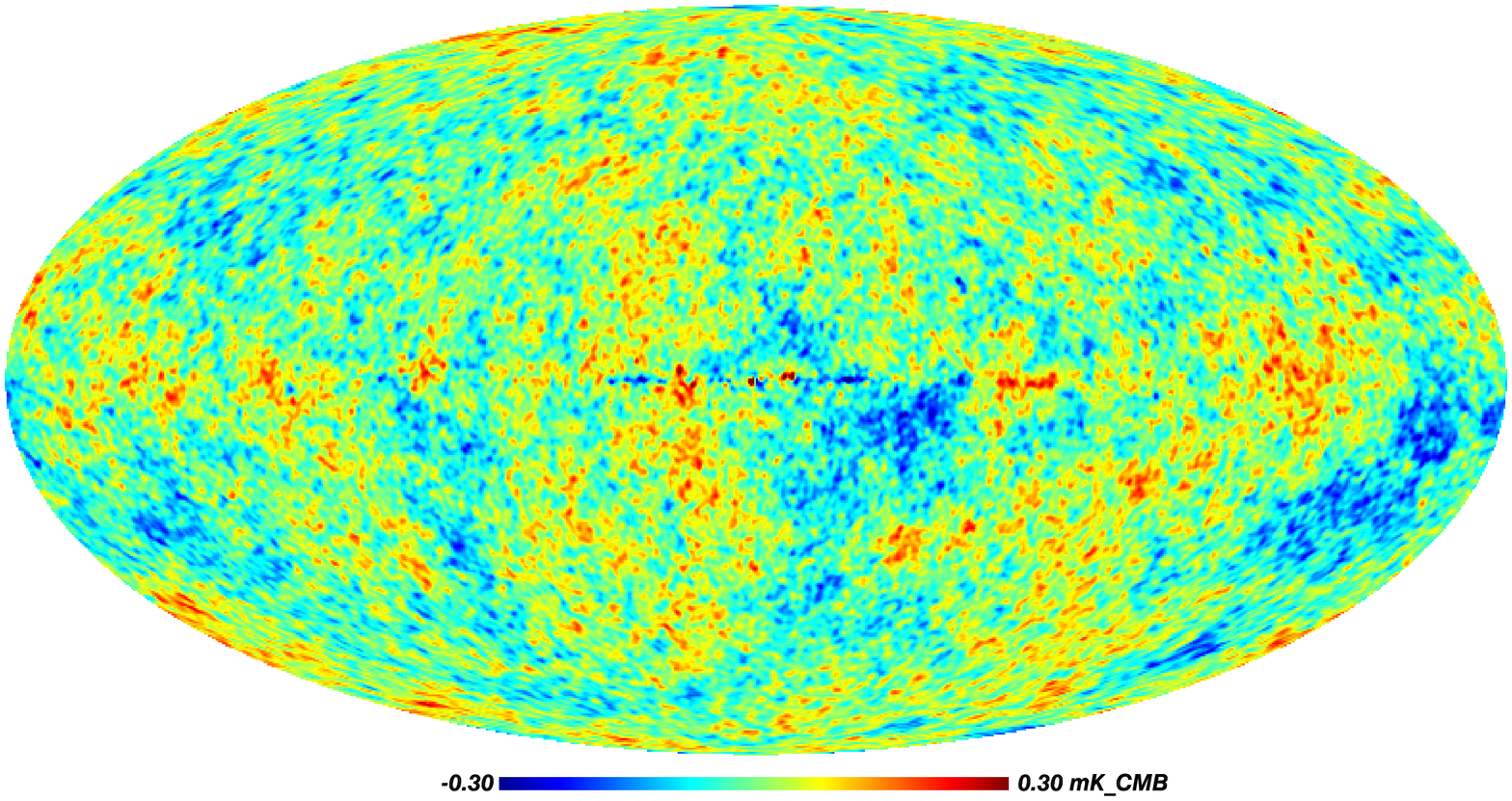}\\
  \includegraphics[scale=0.3,angle=90]{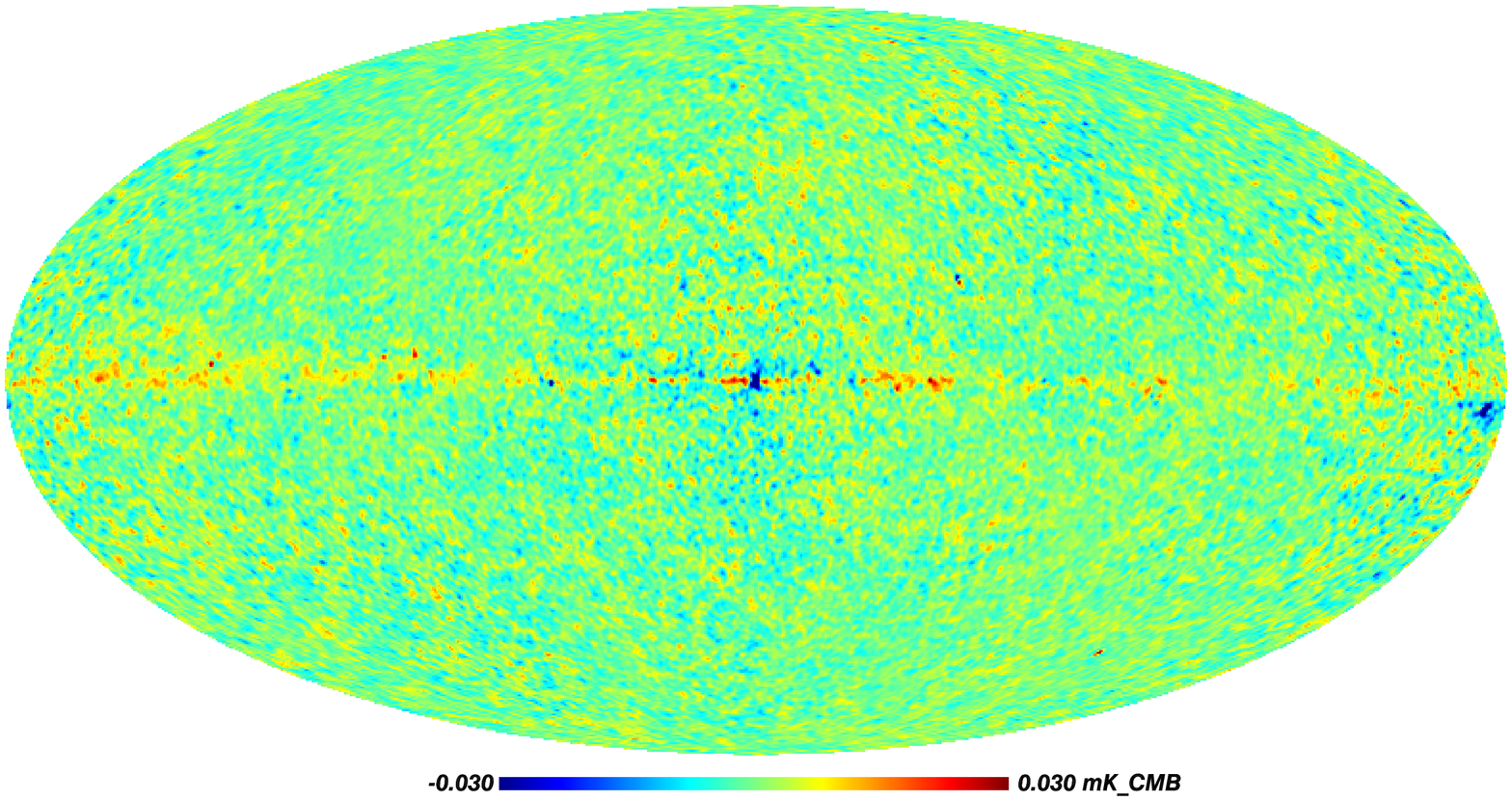}\\
  \includegraphics[scale=0.3,angle=90]{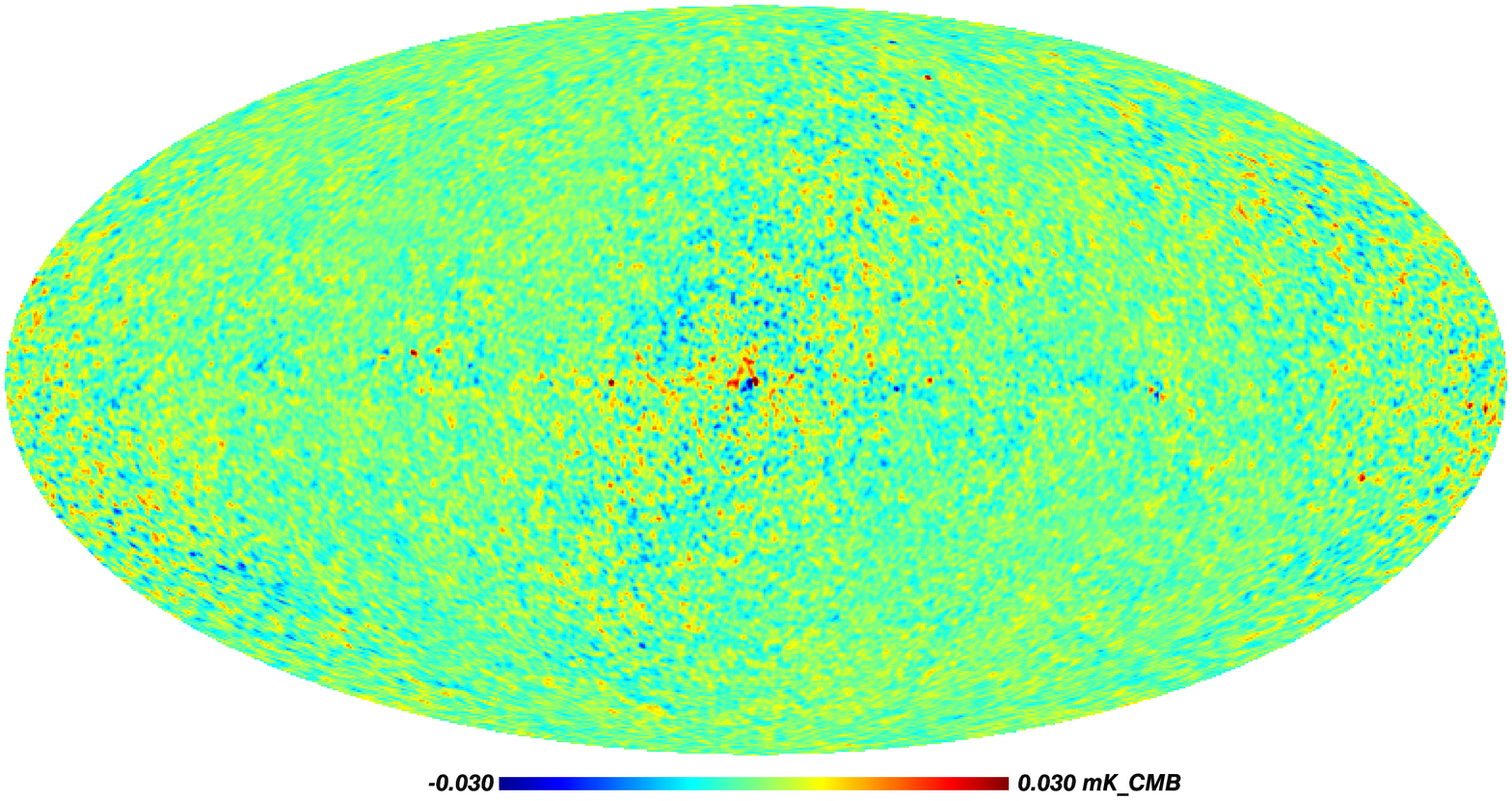}\\
  \caption{The NILC estimate of temperature anisotropies(top) and,
    Stokes parameters $Q$ (middle) and $U$ (bottom) of CMB obtained by
    implementing the NILC on the $9$-year band average maps, at
    nside$~=~512$ and degraded to $60$ arcmin resolution.}
  \label{fig:cmb-map9yr}
\end{figure}
\section{WMAP $9$-year needlet ILC spectrum}
\label{sec:wmap-spec}
As already mentioned above, the foreground-cleaned maps of $T$, $E$
and $B$ obtained in this way are not fully exempt from contamination
by residual foregrounds and noise. The recovered CMB map ($T$, $Q$ and
$U$) at $60$ arcmin resolution is displayed in
figure~\ref{fig:cmb-map9yr}. Residual foreground contamination, albeit
small, is visible along a narrow strip on the Galactic plane of these
maps. Noise contamination is seen from the larger variance of the
polarisation maps away from the ecliptic poles. Note, however, that
these maps have been obtained with no masking whatsoever of the
original data sets. We rely on the localisation provided by the
needlets to avoid more contaminated sky areas to impact the
reconstruction of the CMB over the rest of the sky. That choice, even
if probably sub-optimal, is easy to implement, and turns out to be
good enough for the analysis of WMAP observations.

\subsection{Minimisation of the impact of noise}

Noise biasing in our estimated power spectra is avoided by producing,
for each of $T$, $E$ and $B$, an independent CMB map for each of the
$9$ individual years of observation. All maps, however, are obtained
using the same set of needlet weights, determined using the co-added
$9$ year observations. To avoid the bias induced by residual
instrumental noise in the maps, we compute the CMB power spectra
exclusively from cross-products between maps from different
years. Each data point in our spectra is thus obtained as an average
of all possible cross-year spectra ($72$, although in the case of $EE$
and $BB$ half of those are strictly identical to the other half). We
take into account the correlation between errors in the $72$
cross-spectra estimates for the final average spectrum (see appendix
for details).
%

\subsection{Minimisation of the impact of foregrounds}
Unlike the residual noise, the residual foreground emission in each of
$9$ cleaned maps is the same, as each map observes the same sky
emission, and is produced using the same linear combination for all
years.

We use the conservative temperature\footnote{\small
  http://lambda.gsfc.nasa.gov/data/map/dr5/ancillary/masks/\\wmap\_tempearture\_kq85\_analysis\_mask\_r9\_9yr\_v5.fits}
and polarisation\footnote{\small
  http://lambda.gsfc.nasa.gov/data/map/dr5/ancillary/masks/\\wmap\_polarization\_analysis\_mask\_r9\_9yr\_v5.fits}
analysis masks provided by the WMAP collaboration, applied directly on
the CMB maps after the needlet ILC. We choose this option (rather than
masking before the ILC) so that we produce full sky CMB maps that can
also be used for other purposes than power spectrum estimation. In
order to correct for the sky fraction we use here the MASTER method
\citep{2002ApJ...567....2H} to compute the power spectrum.

\subsection{The effectiveness of the NILC}
The NILC approach automatically adjusts weights of the linear
combination input sky maps as a function of sky area and of angular
scale (i.e. in regions of pixel-scale space), to minimise the overall
contamination by any signal that does not have the expected colour of
the CMB. This includes astrophysical foregrounds, but also
instrumental noise and even residual additive systematic effects.

For instance, in a pixel-scale region where instrumental noise
dominates the error in the observations (i.e. with negligible
foregrounds), the NILC is in effect equivalent to a noise-weighted
average of all WMAP maps. All the ILC weights are positive,
proportional to the inverse of the noise power of the various channels
in that region.

On the other hand, in a pixel-scale region significantly contaminated
by foregrounds, the NILC coefficients adjust themselves automatically
to minimise the total variance of the error, i.e. use positive and
negative coefficients to cancel-out the foregrounds (in a compromise
between noise and foreground contamination).

Finally, the NILC is even effective at minimising the contamination by
unknown additive residual systematics. Imagine that one particular
channel suffers from such residuals in one particular pixel-scale
region. The NILC will automatically minimise the weight of that
particular channel in the linear combination for that region (and,
possibly, the weight of another channel in another region, if that
turns out to be necessary).

\subsection{Impact of calibration errors}
An important assumption of the ILC is that the frequency scaling of
the CMB is known. However, calibration coefficients for each channel,
which are a multiplicative factor for each frequency, introduce an
uncertainty in the frequency scalings of the CMB component in the
presence of calibration errors \citep{2010MNRAS.401.1602D}. This
effect is particularly strong in the high signal to noise ratio
regime. Considering the relatively low signal to noise of WMAP
polarised maps, this issue can safely be ignored here.

Beam uncertainties induce similar biases as calibration uncertainties,
except that these biases are scale dependent. Here again, such biases
are not the main source of error in our final polarisation spectra, as
their impact is small in comparison to the uncertainties due to
instrumental noise for polarisation measurements with WMAP.

These issues, connected to the exact response of the detectors,
however, will require specific attention with upcoming more sensitive
observations of the polarised CMB, if our method is to be used for the
analysis of these future data sets.

\subsection{Noise-weighting}
Residual noise in our CMB maps is inhomogeneous, primarily because of
non-uniform sky coverage, with higher number of observations in the
directions of ecliptic poles and rings at $45^{\circ}$ ecliptic
latitude. 

At multipoles where noise is the main source of error, there is
advantage to weighting the maps with the inverse noise variance for
computing the power spectrum. This amounts to giving more weight in
the final spectrum estimate to regions of the sky less contaminated by
noise. At large angular scales, however, cosmic variance dominates,
and it is preferable to use uniform weighting.

Considering this, we compute all power spectra using both schemes, for
all multipole bins. In practice, the map of weights for the
noise-weighted scheme is obtained using the map of number of hits of
the W-channel (maps for all channels are similar).

For both cases (noise weighted and uniform weighting), we compute the
error bar on our estimate of CMB power spectra as described in the
appendix. We then pick, for the final power spectrum, that of the two
with the lowest variance. In the case of $TE$ and $TB$, the uniform
weighting is better in the first 24 multipole bins (below
$l=317$). For $EE$, $BB$ and $EB$, the noise-weighted estimate is
better in all multipole bins.

An alternative to this noise-weighting method would be to use the
optimised needlet weighting approach investigated for intensity maps
in a method paper by \citet{2008PhRvD..78h3013F}, which pushes the
optimisation yet one step further and is used in our previous analysis
of WMAP intensity maps \citep{2012MNRAS.419.1163B}. However, the extra
complication involved is not necessary here, as it does not make much
difference for a data set in which the noise is not too inhomogeneous
(as is the case in the present data set).

\subsection{Results}
We now present the estimated polarisation power spectra and
temperature-polarisation cross power spectra obtained on the basis of
WMAP $9$-year observations. The error bars in our estimates (see
appendix for details) include the total statistical error of the
estimator (noise and cosmic variance terms). 
The measurement of the variance of CMB power spectra,
requires an estimate of the power spectrum of the esidual noise
present in NILC-CMB maps (see Appendix for details). This estimation
could be made with a blind method such as SMICA \citep{2003MNRAS.346.1089D,2008ISTSP...2..735C}, which provides a 
maximum-likelihood multi-component fit to an empirical estimate of the multi-varied power
spectrum of several independent observations of CMB contaminated by foegrounds and noise.
Here, we perform a simpler estimation in three steps. First, we average all
possible single-year measurements power spectra, corrected for the effect of the mask
using the MASTER method. Then, we estimate the noise level from the
difference of this measurement based on on-diagonal terms, and of the CMB power
spectrum inferred from off-diagonal terms in the maps covariance. 
Finally, we obtain the variance of CMB power spectra
from this together with best-fit theoretical power
spectra, using equation \ref{equ:est_variance}. We correct our error estimates
for partial sky coverage by
dividing the variance of measured angular power spectra by the
corresponding sky
fraction (equation \ref{equ:fskyt} and
  \ref{equ:fskyp}).

\begin{figure}
  \centering
  \includegraphics[scale=0.27,angle=90]{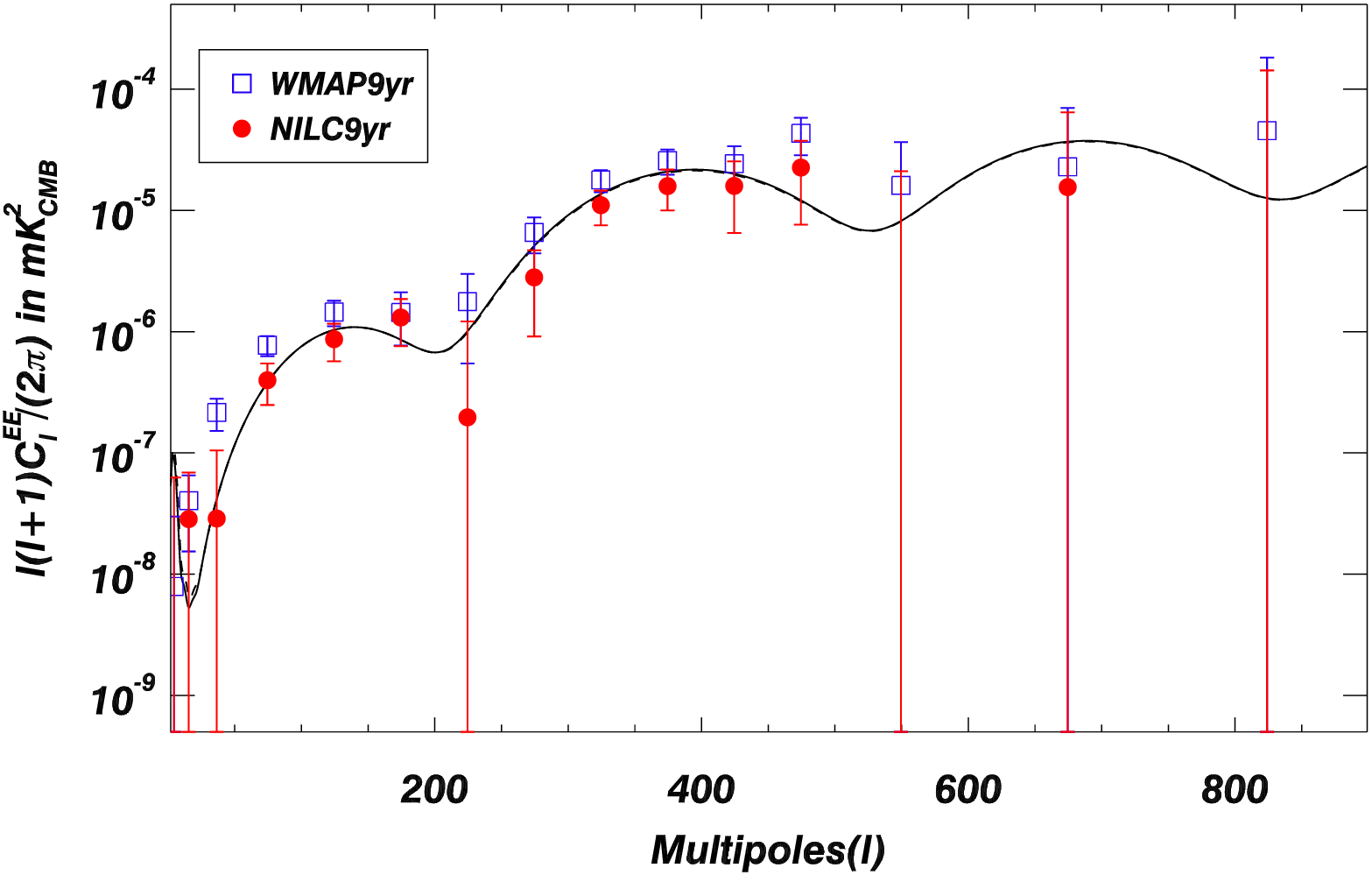}
  \includegraphics[scale=0.27,angle=90]{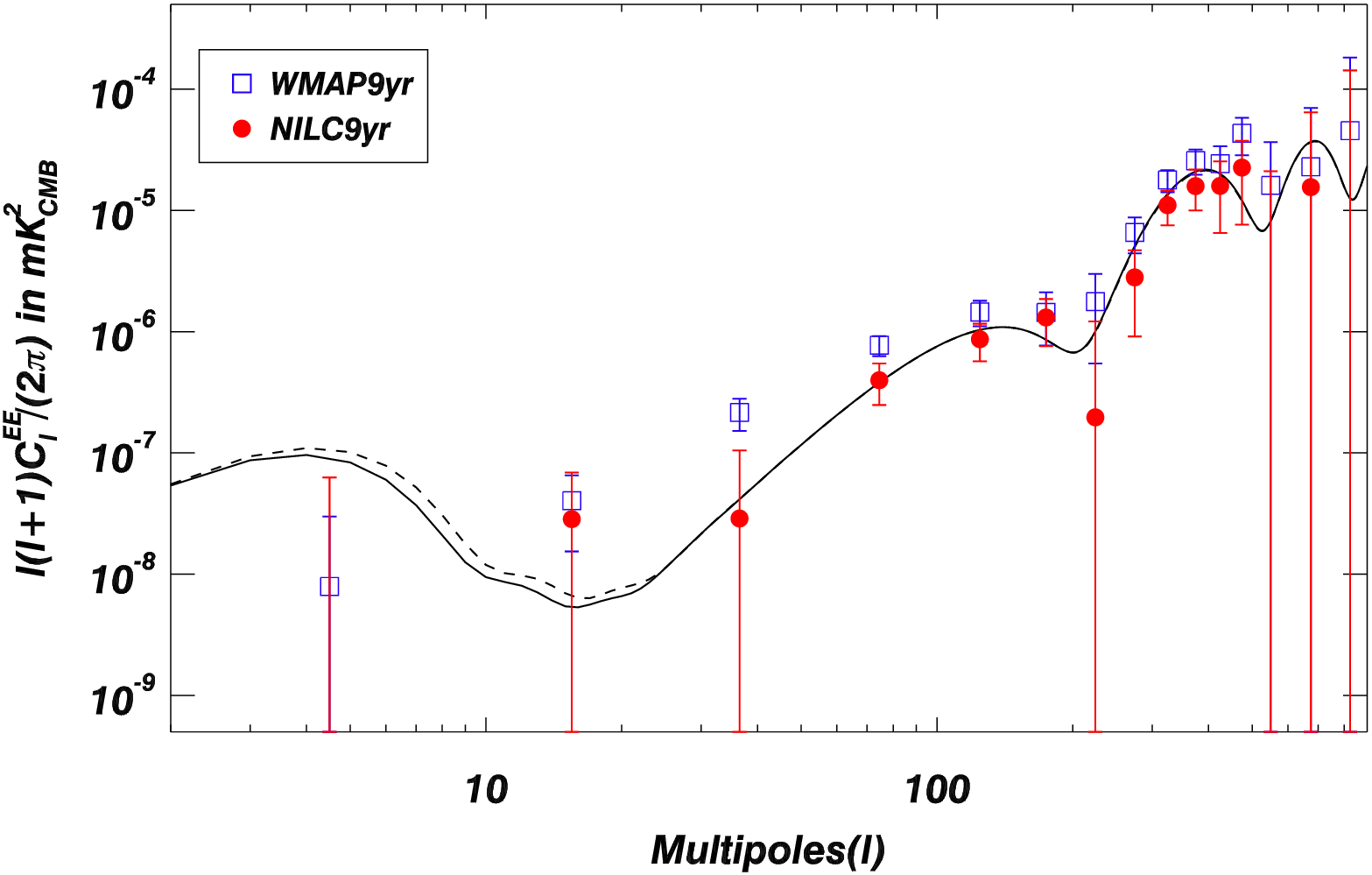}
  \caption{The red filled circles show the angular power spectrum of $E$-mode
    of CMB polarisation as estimated with our method using $9$ years
    of observations of WMAP. The dark blue open squares show the $9$-year
    angular power spectrum of $E$-mode of CMB polarisation published
    by the WMAP collaboration. The solid black line shows the
    theoretical angular power spectrum for WMAP best-fit $\Lambda$-CDM
    model \citep{2012arXiv1212.5226H}. The broken black line shows the
    theoretical angular power spectrum for Planck best-fit
    $\Lambda$-CDM model \citep {2013arXiv1303.5076P}. The top panel
    uses a linear scale in the horizontal axis, and the bottom panel a
    logarithmic scale.}
  \label{fig:cmb-cl-ee}
\end{figure}


\begin{table}
\caption{Comparison of our estimate of binned angular power spectrum
  of $E$-mode of CMB polarisation with that provided by WMAP team. The
  quantities tabulated are $D_l = l(l+1)C_l^{EE}/2\pi$ and $\Delta D_l
  = l(l+1) \Delta C_l^{EE}/2\pi$.}
  \centering \begin{tabular}{c c c c c}
  \hline\hline 
  $l_{\rm range}$ & $D_l^{\rm nilc}$ & $D_l^{\rm wmap}$ & $\Delta D_l^{\rm nilc}$ & $\Delta D_l^{\rm wmap}$ \\ 
  $ $ & $ $ & $ $ & $ $ & $ $\\ 
  $ $ & $(mK^{2})$ & $(mK^{2})$ & $(mK^{2})$ & $(mK^{2})$ \\ 
[1ex] \hline
        2--7  & -1.378e-08  &  7.917e-09  &  6.776e-08  &  2.191e-08 \\ 
       8--23  &  1.323e-08  &  4.035e-08  &  4.700e-08  &  2.496e-08 \\ 
      24--49  &  8.213e-09  &  2.159e-07  &  8.574e-08  &  6.396e-08 \\ 
      50--99  &  3.574e-07  &  7.733e-07  &  1.679e-07  &  1.452e-07 \\ 
    100--149  &  7.732e-07  &  1.453e-06  &  3.329e-07  &  3.457e-07 \\ 
    150--199  &  1.493e-06  &  1.441e-06  &  6.169e-07  &  6.721e-07 \\ 
    200--249  & -3.088e-07  &  1.769e-06  &  1.142e-06  &  1.223e-06 \\ 
    250--299  &  2.289e-06  &  6.583e-06  &  2.111e-06  &  2.160e-06 \\ 
    300--349  &  9.964e-06  &  1.783e-05  &  3.928e-06  &  3.673e-06 \\ 
    350--399  &  1.145e-05  &  2.572e-05  &  6.534e-06  &  6.007e-06 \\ 
    400--449  &  1.326e-05  &  2.431e-05  &  1.050e-05  &  9.510e-06 \\ 
    450--499  &  2.051e-05  &  4.327e-05  &  1.667e-05  &  1.478e-05 \\ 
    500--599  & -4.335e-05  &  1.609e-05  &  2.352e-05  &  2.055e-05 \\ 
    600--749  &  1.814e-05  &  2.289e-05  &  5.443e-05  &  4.718e-05 \\ 
    750--898  & -2.773e-04  &  4.548e-05  &  1.594e-04  &  1.367e-04 \\ 
 [1ex] \hline
\end{tabular} 
\label{tab:binned_eespec} 
\end{table}

\subsubsection{$EE$ angular power spectrum}
Figure \ref{fig:cmb-cl-ee} shows the estimated auto-angular power
spectrum for the $E$-mode of CMB polarisation. Our estimated $E$-mode
CMB power is lower than that obtained by the WMAP
collaboration\footnote{\small
  http://lambda.gsfc.nasa.gov/data/map/dr5/dcp/spectra/\\wmap\_ee\_spectrum\_9yr\_v5.txt},
in all of the multipole bins, and in better agreement with the
theoretical expectations (assuming the cosmological model is correct)
(see table \ref{tab:binned_eespec}). The systematic difference between
our measurement and that of the WMAP team is presently not
understood. We suspect that their estimate is contaminated by residual
foreground emission or systematics.

\begin{figure}
  \centering
  \includegraphics[scale=0.27,angle=90]{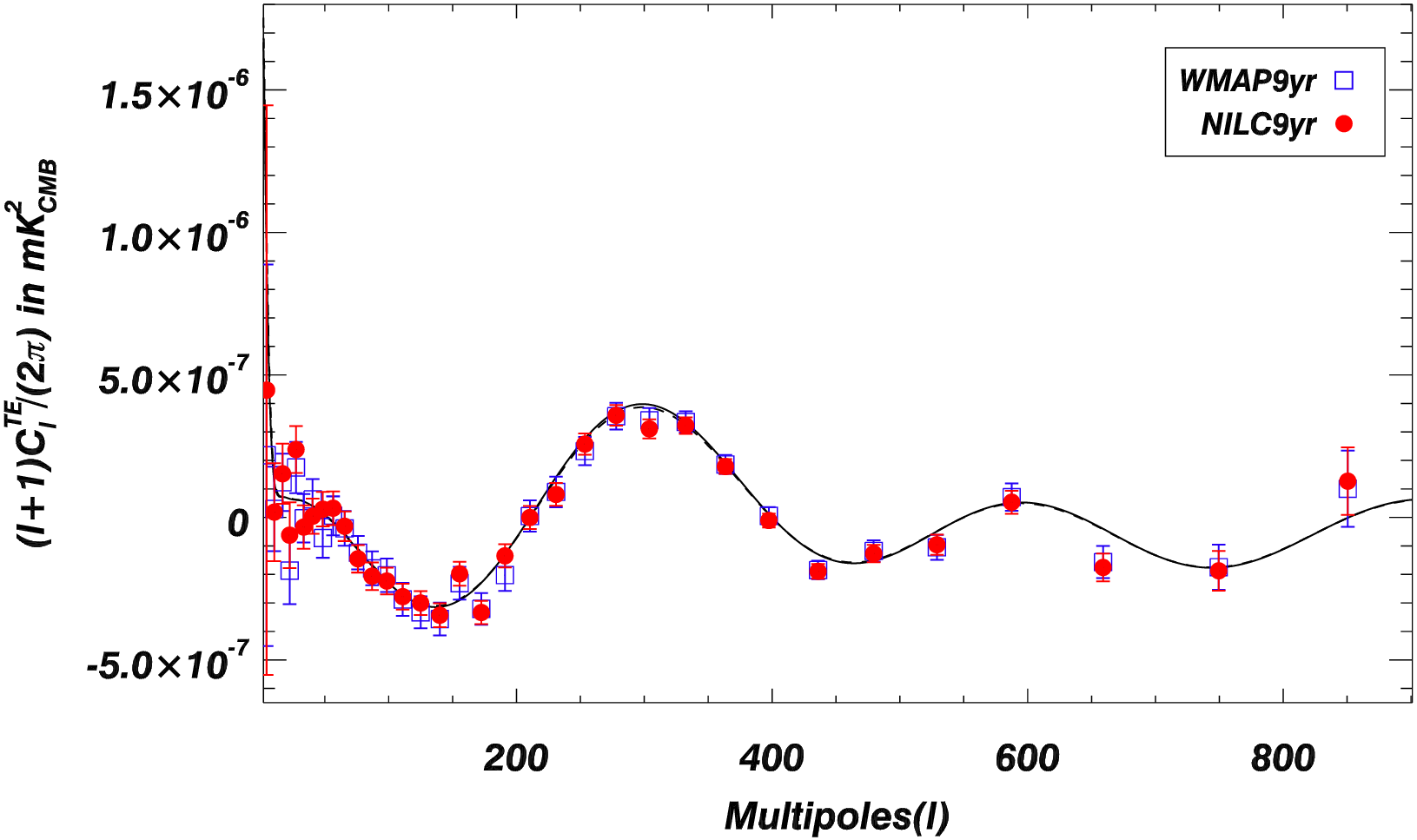}
  \includegraphics[scale=0.27,angle=90]{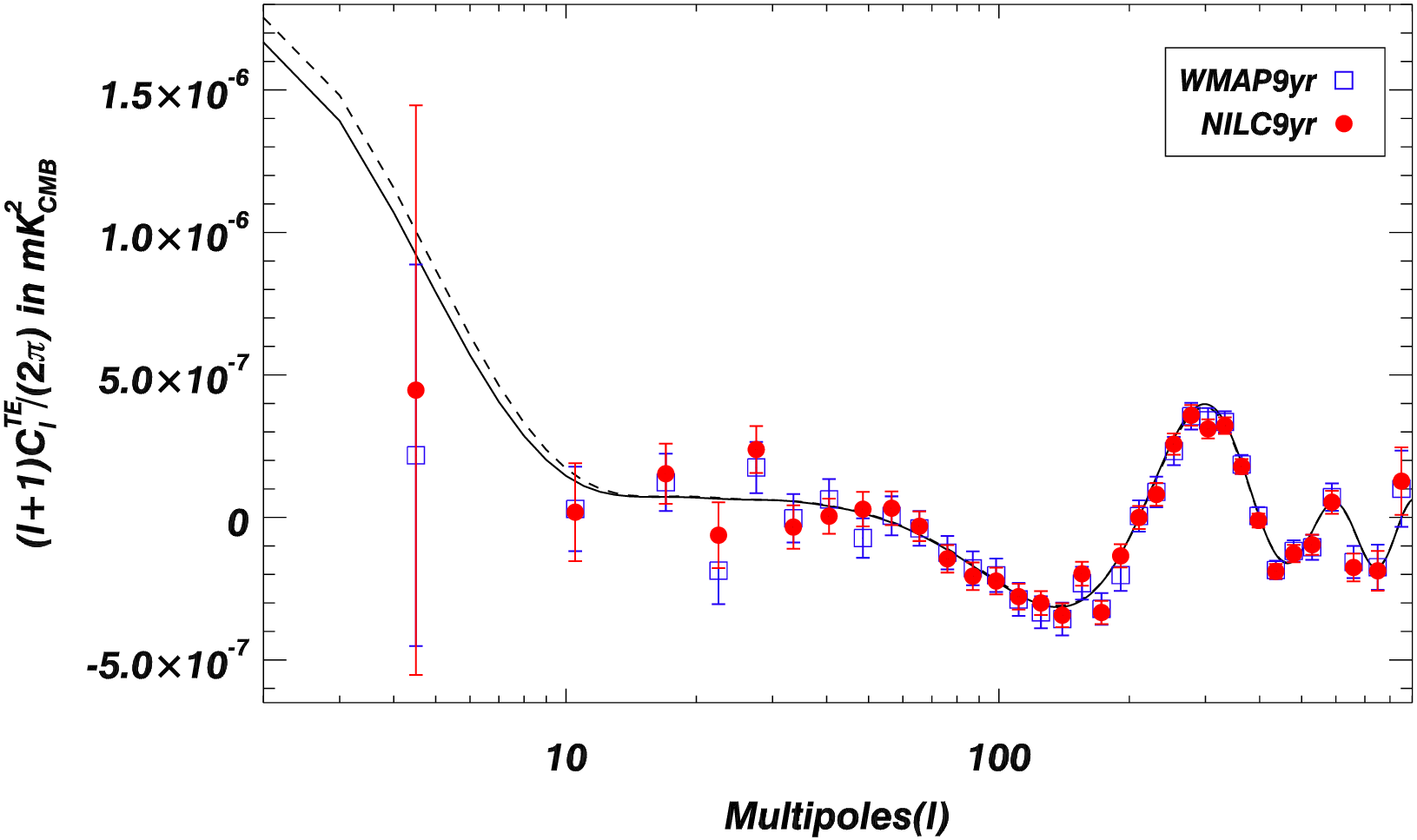}
  \caption{The red filled circles show the cross-angular power spectrum of
    CMB temperature anisotropies ($T$-mode) and $E$-mode of CMB
    polarisation as estimated with our method using $9$ years of
    observation of WMAP. The dark blue open squares show the $9$-year
    angular power spectrum of $E$-mode of CMB polarisation published
    by the WMAP collaboration. The black solid line shows the
    theoretical angular power spectrum for WMAP best-fit $\Lambda$-CDM
    model \citep{2012arXiv1212.5226H}. The broken black line shows the
    theoretical angular power spectrum for Planck best-fit
    $\Lambda$-CDM model \citep {2013arXiv1303.5076P}. The top panel
    uses a linear scale in the horizontal axis, and the bottom panel a
    logarithmic scale.}
  \label{fig:cmb-cl-te}
\end{figure}
\begin{table}
\caption{Comparison of our estimate of binned cross-angular power
  spectrum of CMB temperature anisotropies ($T$-mode) and $E$-mode of
  CMB polarisation with that provided by WMAP team. The quantities
  tabulated are $D^{\prime}_l = (l+1)C_l^{TE}/2\pi$ and $\Delta
  D^{\prime}_l = (l+1)\Delta C_l^{TE}/2\pi$.}
  \centering \begin{tabular}{c c c c c} \hline\hline 
  $l_{\rm range}$ & $D_l^{\prime,\rm nilc}$ & $D_l^{\prime,\rm wmap}$ & $\Delta D_l^{\prime,\rm nilc}$ & $\Delta D_l^{\prime,\rm wmap}$\\ 
  $ $ & $ $ & $ $ & $ $ & $ $ \\ 
  $ $ & (mK$^{2}$) & (mK$^{2}$) & (mK$^{2}$) & (mK$^{2}$) \\ 
[1ex]\hline
        2--7  &  4.467e-07  &  2.181e-07  &  9.996e-07  &  6.696e-07 \\ 
       8--13  &  1.855e-08  &  2.980e-08  &  1.718e-07  &  1.479e-07 \\ 
      14--20  &  1.533e-07  &  1.233e-07  &  1.058e-07  &  1.007e-07 \\ 
      21--24  & -6.235e-08  & -1.866e-07  &  1.152e-07  &  1.178e-07 \\ 
      25--30  &  2.382e-07  &  1.752e-07  &  8.256e-08  &  8.999e-08 \\ 
      31--36  & -3.389e-08  & -2.881e-09  &  7.634e-08  &  8.525e-08 \\ 
      37--44  &  4.197e-09  &  6.334e-08  &  6.175e-08  &  7.097e-08 \\ 
      45--52  &  2.877e-08  & -7.248e-08  &  6.060e-08  &  6.912e-08 \\ 
      53--60  &  3.225e-08  &  5.807e-09  &  5.909e-08  &  6.827e-08 \\ 
      61--70  & -3.099e-08  & -3.864e-08  &  5.205e-08  &  6.094e-08 \\ 
      71--81  & -1.452e-07  & -1.238e-07  &  4.829e-08  &  5.862e-08 \\ 
      82--92  & -2.062e-07  & -1.786e-07  &  4.841e-08  &  5.968e-08 \\ 
     93--104  & -2.232e-07  & -2.030e-07  &  4.676e-08  &  5.854e-08 \\ 
    105--117  & -2.780e-07  & -2.879e-07  &  4.556e-08  &  5.792e-08 \\ 
    118--132  & -3.003e-07  & -3.328e-07  &  4.197e-08  &  5.575e-08 \\ 
    133--147  & -3.431e-07  & -3.566e-07  &  4.200e-08  &  5.761e-08 \\ 
    148--163  & -1.978e-07  & -2.309e-07  &  4.196e-08  &  5.744e-08 \\ 
    164--181  & -3.337e-07  & -3.209e-07  &  4.079e-08  &  5.562e-08 \\ 
    182--200  & -1.348e-07  & -2.025e-07  &  4.053e-08  &  5.535e-08 \\ 
    201--220  & -3.575e-10  &  5.030e-09  &  4.022e-08  &  5.475e-08 \\ 
    221--241  &  8.056e-08  &  8.948e-08  &  3.998e-08  &  5.367e-08 \\ 
    242--265  &  2.573e-07  &  2.327e-07  &  3.741e-08  &  4.966e-08 \\ 
    266--290  &  3.583e-07  &  3.552e-07  &  3.599e-08  &  4.706e-08 \\ 
    291--317  &  3.106e-07  &  3.408e-07  &  3.370e-08  &  4.264e-08 \\ 
    318--347  &  3.389e-07  &  3.350e-07  &  3.030e-08  &  3.715e-08 \\ 
    348--379  &  1.787e-07  &  1.859e-07  &  2.758e-08  &  3.311e-08 \\ 
    380--415  & -1.111e-08  &  5.304e-09  &  2.605e-08  &  3.075e-08 \\ 
    416--456  & -2.042e-07  & -1.845e-07  &  2.794e-08  &  3.224e-08 \\ 
    457--502  & -1.151e-07  & -1.176e-07  &  3.268e-08  &  3.717e-08 \\ 
    503--555  & -1.018e-07  & -1.049e-07  &  3.818e-08  &  4.379e-08 \\ 
    556--619  &  5.537e-08  &  7.127e-08  &  4.356e-08  &  4.780e-08 \\ 
    620--698  & -1.882e-07  & -1.567e-07  &  5.307e-08  &  5.623e-08 \\ 
    699--800  & -2.007e-07  & -1.747e-07  &  7.632e-08  &  7.940e-08 \\ 
    801--900  &  1.935e-07  &  1.008e-07  &  1.299e-07  &  1.338e-07 \\ 
 [1ex] \hline
\end{tabular} 
\label{tab:binned_tespec} 
\end{table}
\subsubsection{$TE$ cross-spectrum}
Figure \ref{fig:cmb-cl-te} shows the estimated cross-angular power
spectrum for temperature anisotropy of CMB and $E$-mode of CMB
polarisation. The power spectrum obtained using our analysis is in good agreement
 with that provided by the WMAP collaboration\footnote{\small
  http://lambda.gsfc.nasa.gov/data/map/dr5/dcp/spectra/\\wmap\_te\_spectrum\_9yr\_v5.txt}
and with the WMAP best-fit
model (see table \ref{tab:binned_tespec}).
\begin{figure}
  \centering
  \includegraphics[scale=0.27,angle=90]{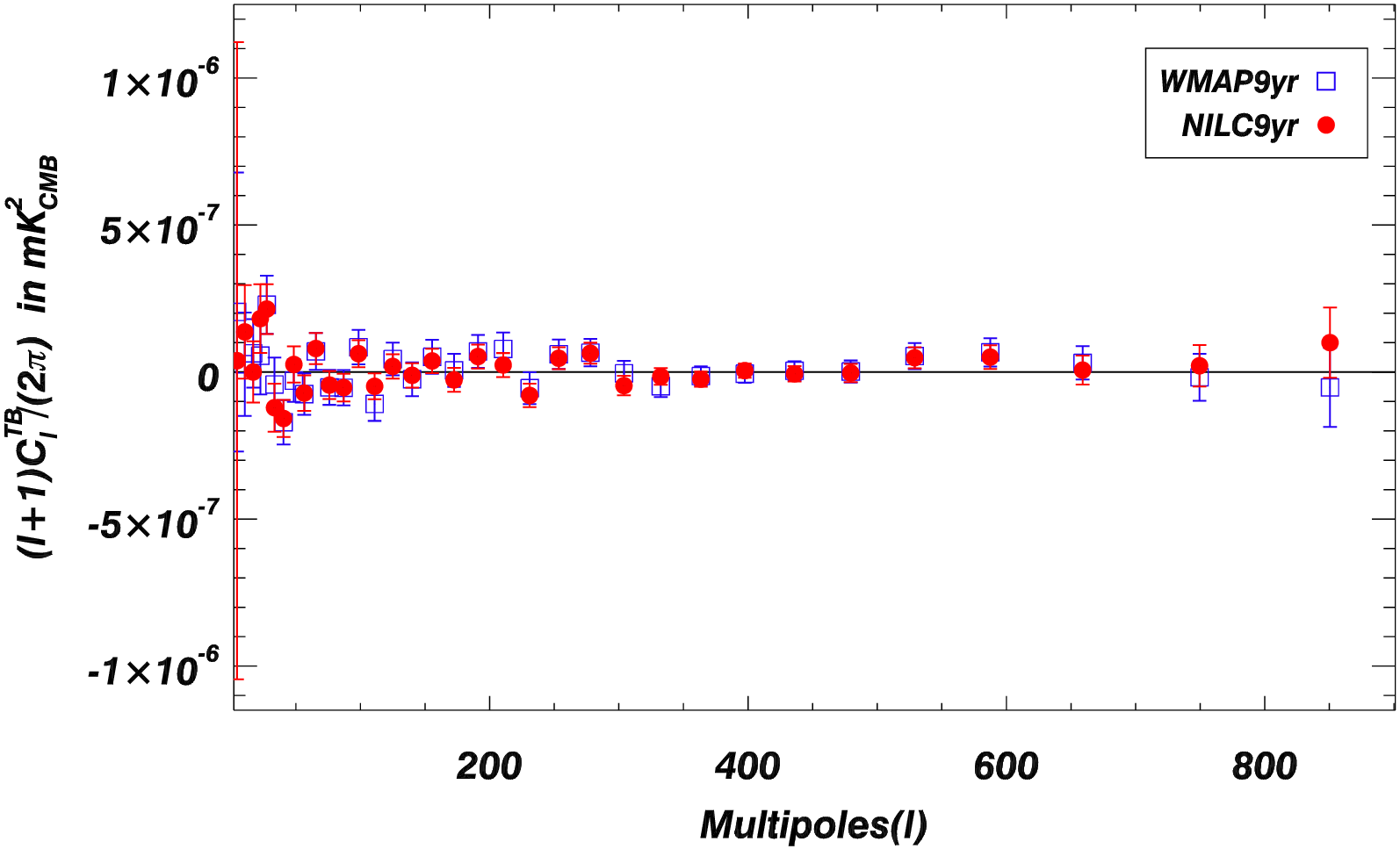}
  \includegraphics[scale=0.27,angle=90]{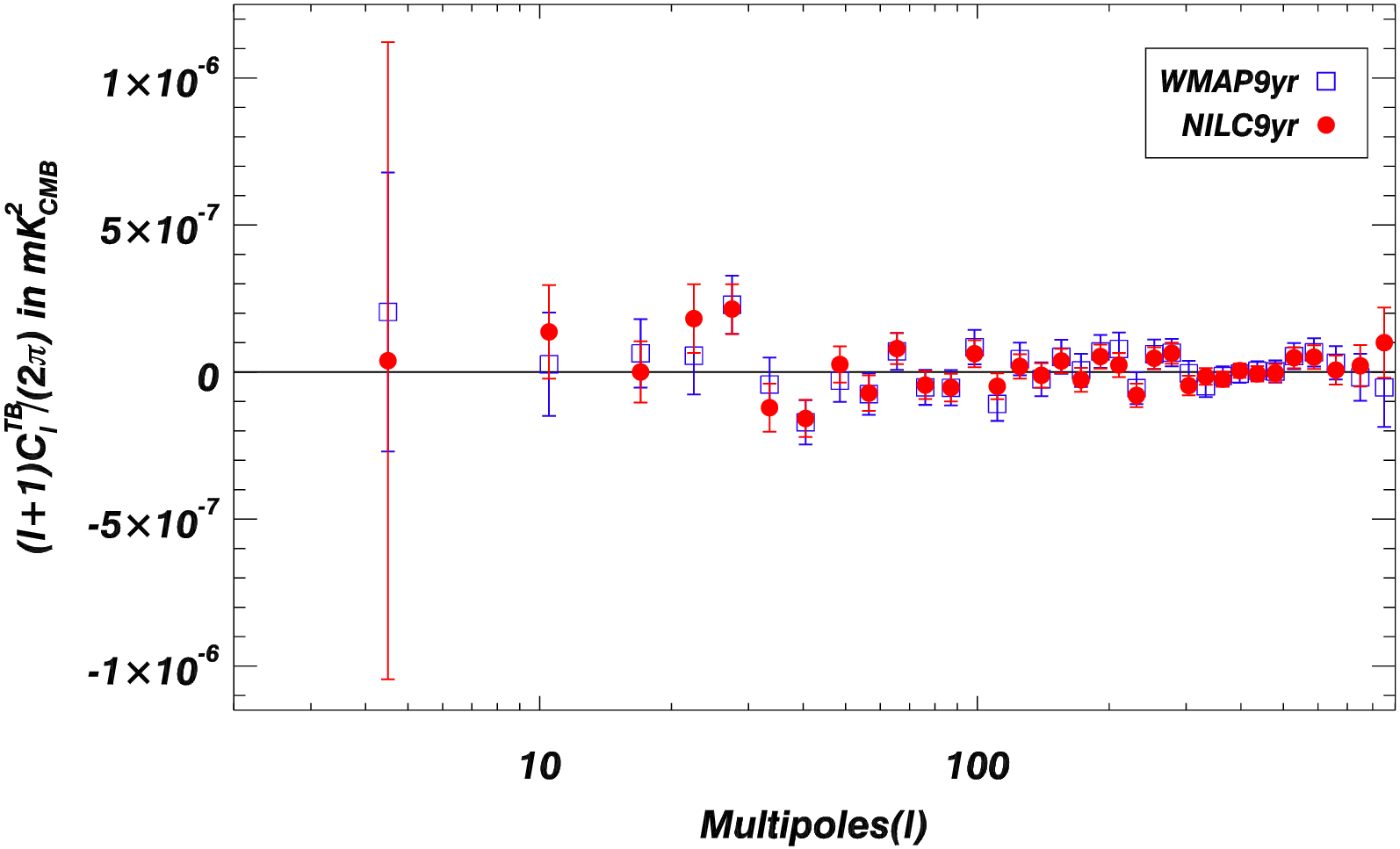}
  \caption{The red filled circles show the cross-angular power spectrum of
    CMB temperature anisotropies ($T$-mode) and $B$-mode of CMB
    polarisation estimated using $9$ years of observation of WMAP. The
    dark blue open squares show the $9$-year angular power spectrum of
    $E$-mode of CMB polarisation published by the WMAP
    collaboration. The top panel uses a linear scale in the horizontal
    axis, and the bottom panel a logarithmic scale.}
  \label{fig:cmb-cl-tb}
\end{figure}
\begin{table}
\caption{Comparison of our estimate of binned cross-angular power
  spectrum of CMB temperature anisotropies ($T$-mode) and $B$-mode of CMB
  polarisation with that provided by WMAP team. The quantities
  tabulated are $D^{\prime}_l = (l+1)C_l^{TB}/2\pi$ and $\Delta
  D^{\prime}_l = (l+1)\Delta C_l^{TB}/2\pi$.}  
  \centering \begin{tabular}{c c c c c} \hline\hline 
  $l_{\rm range}$ & $D_l^{\prime,\rm nilc}$ & $D_l^{\prime,\rm wmap}$ & $\Delta D_l^{\prime,\rm nilc}$ & $\Delta D_l^{\prime,\rm wmap}$\\ 
  $ $ & $ $ & $ $ & $ $ & $ $ \\ 
  $ $ & (mK$^{2}$) & (mK$^{2}$) & (mK$^{2}$) & (mK$^{2}$) \\ 
[1ex]\hline
        2--7  &  3.853e-08  &  2.043e-07  &  1.084e-06  &  4.745e-07 \\ 
       8--13  &  1.366e-07  &  2.655e-08  &  1.590e-07  &  1.757e-07 \\ 
      14--20  &  1.540e-10  &  6.373e-08  &  1.041e-07  &  1.163e-07 \\ 
      21--24  &  1.817e-07  &  5.569e-08  &  1.169e-07  &  1.320e-07 \\ 
      25--30  &  2.142e-07  &  2.289e-07  &  8.465e-08  &  9.887e-08 \\ 
      31--36  & -1.212e-07  & -4.232e-08  &  8.199e-08  &  9.198e-08 \\ 
      37--44  & -1.580e-07  & -1.707e-07  &  6.306e-08  &  7.538e-08 \\ 
      45--52  &  2.576e-08  & -2.907e-08  &  6.179e-08  &  7.239e-08 \\ 
      53--60  & -7.157e-08  & -7.499e-08  &  6.029e-08  &  7.070e-08 \\ 
      61--70  &  8.001e-08  &  7.028e-08  &  5.305e-08  &  6.245e-08 \\ 
      71--81  & -4.424e-08  & -5.215e-08  &  4.764e-08  &  5.947e-08 \\ 
      82--92  & -5.287e-08  & -5.369e-08  &  4.680e-08  &  6.005e-08 \\ 
     93--104  &  6.241e-08  &  8.456e-08  &  4.580e-08  &  5.856e-08 \\ 
    105--117  & -4.843e-08  & -1.083e-07  &  4.419e-08  &  5.776e-08 \\ 
    118--132  &  1.908e-08  &  4.423e-08  &  4.159e-08  &  5.557e-08 \\ 
    133--147  & -1.114e-08  & -2.427e-08  &  4.142e-08  &  5.759e-08 \\ 
    148--163  &  3.772e-08  &  5.198e-08  &  4.159e-08  &  5.768e-08 \\ 
    164--181  & -2.657e-08  &  5.853e-09  &  4.064e-08  &  5.613e-08 \\ 
    182--200  &  5.280e-08  &  6.994e-08  &  4.071e-08  &  5.604e-08 \\ 
    201--220  &  2.395e-08  &  7.868e-08  &  4.072e-08  &  5.545e-08 \\ 
    221--241  & -7.953e-08  & -5.467e-08  &  4.005e-08  &  5.419e-08 \\ 
    242--265  &  4.689e-08  &  6.071e-08  &  3.742e-08  &  4.983e-08 \\ 
    266--290  &  6.394e-08  &  6.602e-08  &  3.576e-08  &  4.688e-08 \\ 
    291--317  & -4.613e-08  & -4.412e-09  &  3.317e-08  &  4.225e-08 \\ 
    318--347  & -1.451e-08  & -4.799e-08  &  3.001e-08  &  3.677e-08 \\ 
    348--379  & -2.619e-08  & -1.354e-08  &  2.727e-08  &  3.286e-08 \\ 
    380--415  & -2.188e-09  & -5.634e-09  &  2.593e-08  &  3.067e-08 \\ 
    416--456  & -9.703e-09  &  4.405e-09  &  2.803e-08  &  3.229e-08 \\ 
    457--502  & -5.652e-10  &  1.875e-09  &  3.297e-08  &  3.737e-08 \\ 
    503--555  &  4.714e-08  &  5.426e-08  &  3.859e-08  &  4.410e-08 \\ 
    556--619  &  6.403e-08  &  6.695e-08  &  4.373e-08  &  4.811e-08 \\ 
    620--698  &  3.229e-09  &  3.128e-08  &  5.358e-08  &  5.655e-08 \\ 
    699--800  &  5.605e-08  & -1.781e-08  &  7.684e-08  &  7.991e-08 \\ 
    801--900  &  8.951e-08  & -5.222e-08  &  1.308e-07  &  1.347e-07 \\ 
[1ex] \hline
\end{tabular} 
\label{tab:binned_tbspec} 
\end{table}
\subsubsection{$TB$ cross-spectrum}
Our result for the $TB$ cross-spectrum, compared to the WMAP
collaboration one\footnote{\small
  http://lambda.gsfc.nasa.gov/data/map/dr5/dcp/spectra/\\wmap\_tb\_spectrum\_9yr\_v5.txt},
is shown in figure~\ref{fig:cmb-cl-tb}. Theoretically, this
cross-spectrum is supposed to vanish (to preserve the parity
symmetry), and significant departure from zero would be the sign of
either unknown systematics in the measurement (including residual
foregrounds), or new physics. Our measurement is indeed compatible
with zero (see table \ref{tab:binned_tbspec}).

\begin{figure}
  \centering
  \includegraphics[scale=0.27,angle=90]{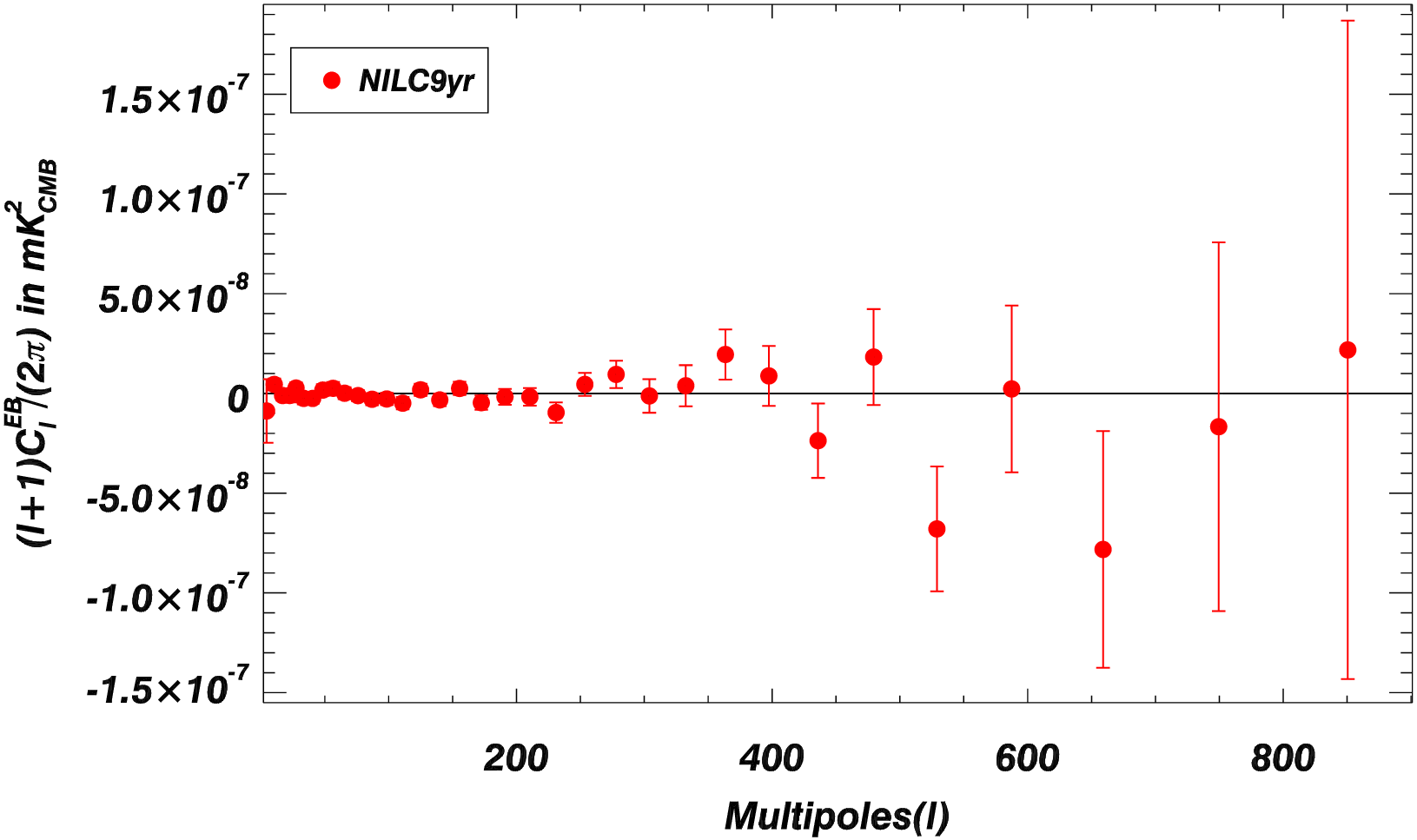}
  \includegraphics[scale=0.27,angle=90]{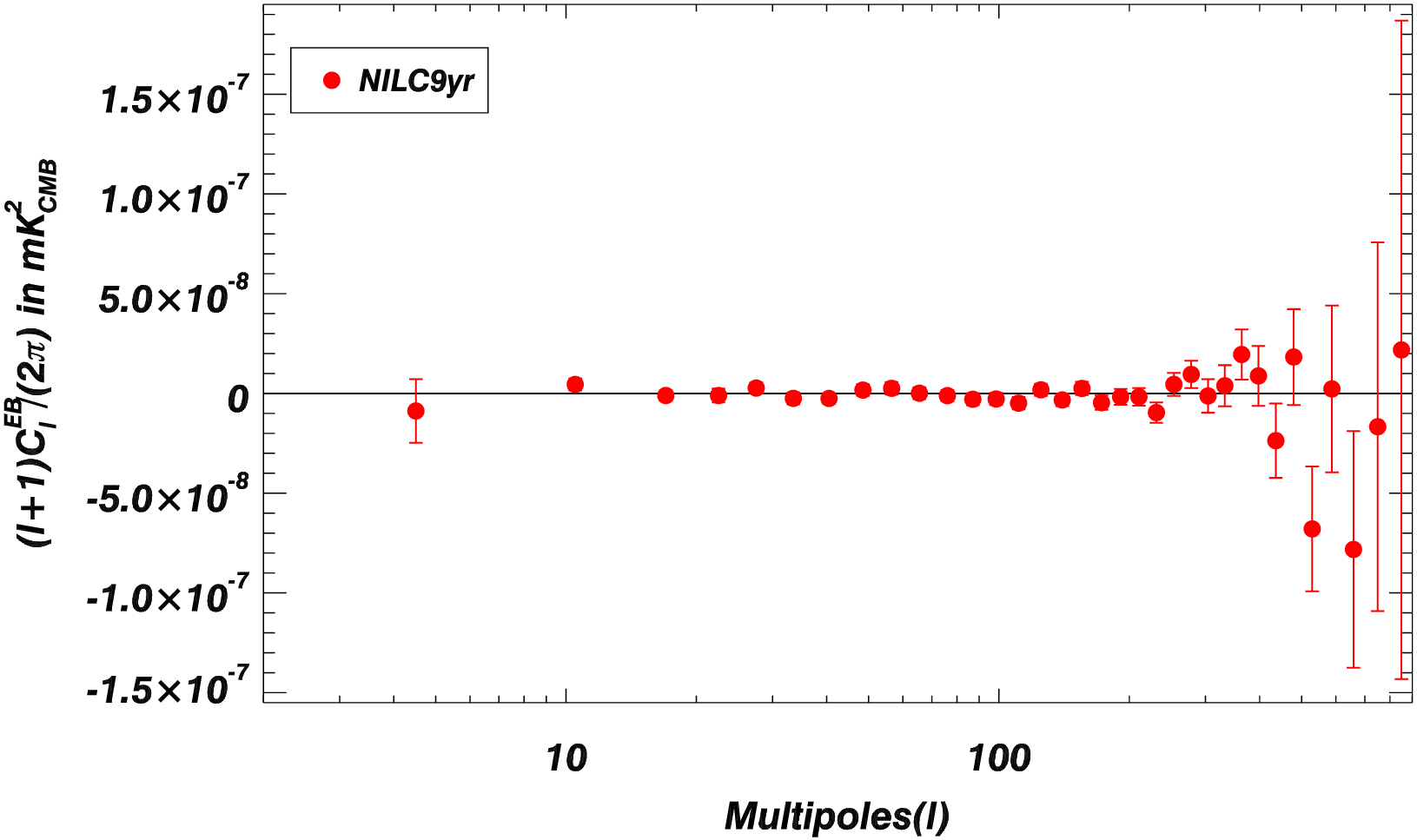}
  \caption{The red filled circles show the cross-angular power spectrum of
    $E$-mode and $B$-mode of CMB polarisation estimated using $9$
    years of observations of WMAP. The top panel uses a linear scale in the
    horizontal axis, and the bottom panel a logarithmic scale. We use
    the same legend and color code for data points in this figure as
    in Figs.~\ref{fig:cmb-cl-te}, \ref{fig:cmb-cl-tb} and
    \ref{fig:cmb-cl-bb}, although WMAP data points are absent here.}
  \label{fig:cmb-cl-eb}
\end{figure}
\begin{table}
\caption{Our estimate of binned cross-angular power spectrum of $E$-mode
  and $B$-mode of CMB polarisation. The quantities
  tabulated are $D^{\prime}_l = (l+1)C_l^{EB}/2\pi$ and $\Delta
  D^{\prime}_l = (l+1)\Delta C_l^{EB}/2\pi$.}
\centering \begin{tabular}{c c c} \hline\hline 
  $l_{\rm range}$ & $D_l^{\prime,\rm nilc}$ & $\Delta D_l^{\prime,\rm nilc}$ \\ $ $ & $ $ & $ $ \\ 
  $ $ & (mK$^{2}$) & (mK$^{2}$)  \\ 
[1ex] \hline
        2--7  & -7.996e-09   &  1.801e-08  \\ 
       8--13  &  5.526e-09   &  3.435e-09  \\ 
      14--20  & -2.265e-09   &  2.968e-09  \\ 
      21--24  & -1.599e-09   &  3.963e-09  \\ 
      25--30  &  3.589e-09   &  3.144e-09  \\ 
      31--36  & -4.133e-09   &  3.442e-09  \\ 
      37--44  & -3.657e-09   &  2.950e-09  \\ 
      45--52  &  1.986e-09   &  3.303e-09  \\ 
      53--60  &  2.373e-09   &  3.512e-09  \\ 
      61--70  & -5.980e-10   &  3.348e-09  \\ 
      71--81  & -3.026e-09   &  3.192e-09  \\ 
      82--92  & -6.053e-09   &  3.350e-09  \\ 
     93--104  & -3.207e-09   &  3.465e-09  \\ 
    105--117  & -6.120e-09   &  3.509e-09  \\ 
    118--132  &  3.808e-09   &  3.380e-09  \\ 
    133--147  & -5.027e-09   &  3.490e-09  \\ 
    148--163  &  1.979e-09   &  3.771e-09  \\ 
    164--181  & -5.674e-09   &  4.013e-09  \\ 
    182--200  & -3.703e-09   &  4.435e-09  \\ 
    201--220  & -3.137e-09   &  5.006e-09  \\ 
    221--241  & -1.110e-08   &  5.731e-09  \\ 
    242--265  &  6.418e-09   &  6.430e-09  \\ 
    266--290  &  1.163e-08   &  7.682e-09  \\ 
    291--317  & -1.343e-09   &  9.405e-09  \\ 
    318--347  &  8.437e-09   &  1.154e-08  \\ 
    348--379  &  2.162e-08   &  1.404e-08  \\ 
    380--415  &  6.150e-09   &  1.677e-08  \\ 
    416--456  & -3.007e-08   &  2.081e-08  \\ 
    457--502  &  2.750e-08   &  2.683e-08  \\ 
    503--555  & -7.581e-08   &  3.500e-08  \\ 
    556--619  &  3.258e-09   &  4.669e-08  \\ 
    620--698  & -1.062e-07   &  6.620e-08  \\ 
    699--800  &  1.443e-09   &  1.031e-07  \\ 
    801--900  & -1.097e-07   &  1.842e-07  \\ 
[1ex] \hline
\end{tabular} 
\label{tab:binned_ebspec} 
\end{table}
\subsubsection{$EB$ cross-spectrum}
The WMAP collaboration has not provided cross-angular power spectra
for $E$ and $B$-modes of CMB polarisation. Our estimated $EB$ cross-power
spectrum, displayed in figure \ref{fig:cmb-cl-eb} and tabulated in
table~\ref{tab:binned_ebspec}, is compatible with zero as expected
from the current cosmological best-fit model.

\begin{figure}
  \centering
  \includegraphics[scale=0.27,angle=90]{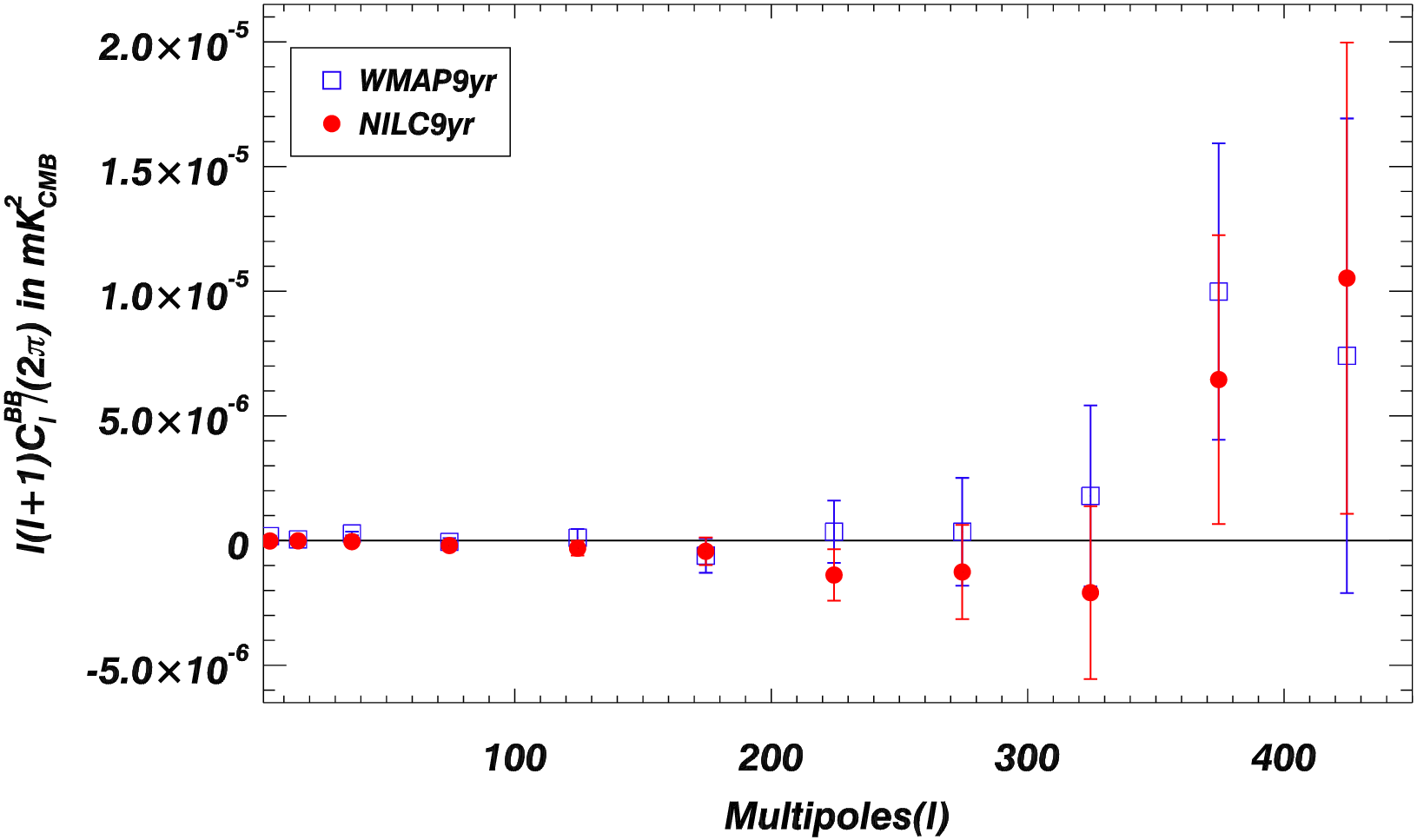}
  \includegraphics[scale=0.27,angle=90]{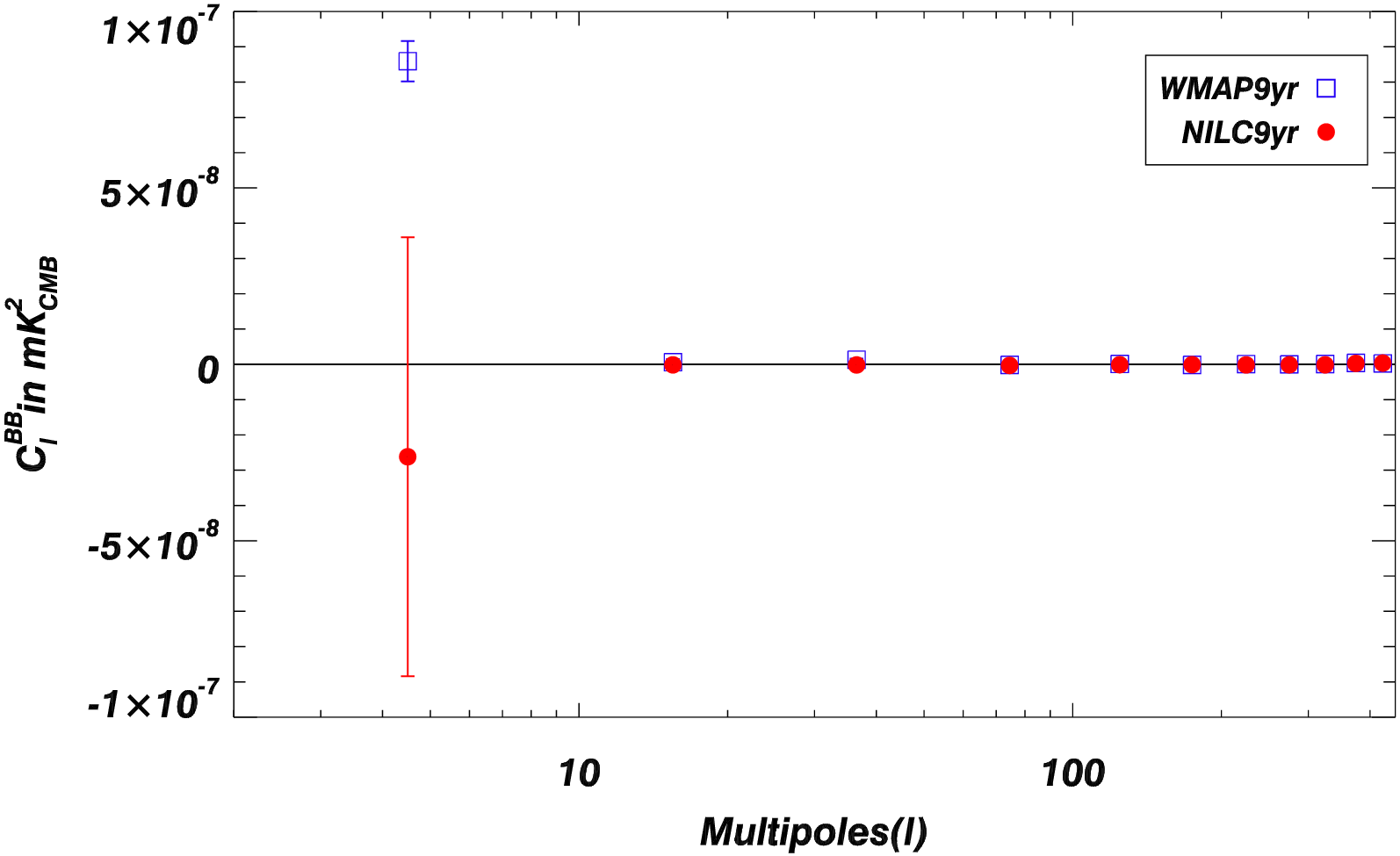}
  \caption{The red filled circles show the angular power spectrum of $B$-mode
    of CMB polarisation estimated using $9$ years of observations of
    WMAP. The top panel uses a linear scale in the horizontal axis,
    and the bottom panels a logarithmic scale. The dark
    blue open squares show the $9$-year angular power spectrum of $B$-mode
    of CMB polarisation published by the WMAP collaboration.}
  \label{fig:cmb-cl-bb}
\end{figure}
\begin{table}
\caption{Comparison of our estimate of binned angular power spectrum
  of $B$-mode of CMB polarisation, with that provided by WMAP team. 
  The quantities tabulated are $D_l = l(l+1)C_l^{BB}/2\pi$ and 
  $\Delta D_l = l(l+1)\Delta C_l^{BB}/2\pi$.}  
  \centering \begin{tabular}{c c c c c c}
  \hline\hline 
  $l_{\rm range}$ & $D_l^{\rm nilc}$ & $D_l^{\rm wmap}$ & $\Delta D_l^{\rm nilc}$ & $\Delta D_l^{\rm wmap}$ \\ 
  $ $ & $ $ & $ $ & $ $ & $ $ \\
  $ $ & (mK$^{2}$) & (mK$^{2}$) & (mK$^{2}$) & (mK$^{2}$) \\
  [1ex]\hline 
        2--7  & -3.748e-08  &  1.776e-07  &  1.178e-07  &  1.683e-08 \\ 
       8--23  & -2.005e-09  &  4.749e-08  &  4.578e-08  &  3.290e-08 \\ 
      24--49  & -2.914e-08  &  2.867e-07  &  9.191e-08  &  7.243e-08 \\ 
      50--99  & -2.457e-07  & -5.148e-08  &  1.647e-07  &  1.494e-07 \\ 
    100--149  & -4.836e-07  &  1.167e-07  &  3.302e-07  &  3.500e-07 \\ 
    150--199  & -2.687e-07  & -6.025e-07  &  6.234e-07  &  6.885e-07 \\ 
    200--249  & -1.804e-06  &  3.528e-07  &  1.160e-06  &  1.248e-06 \\ 
    250--299  & -1.021e-06  &  3.532e-07  &  2.115e-06  &  2.162e-06 \\ 
    300--349  & -2.692e-06  &  1.793e-06  &  3.883e-06  &  3.627e-06 \\ 
    350--399  &  8.445e-06  &  9.989e-06  &  6.468e-06  &  5.946e-06 \\ 
    400--449  &  1.545e-05  &  7.411e-06  &  1.056e-05  &  9.517e-06 \\ 
[1ex] \hline
\end{tabular} 
\label{tab:binned_bbspec} 
\end{table}
\subsubsection{$BB$ angular power spectrum}
Finally, the power spectrum of CMB $B$-modes is of special interest,
as it provides one of the most promising means of detecting primordial
tensor modes in the early universe, and constrain models of
inflation. Significant effort is currently undertaken to prepare the
measurement of $B$-modes with a future space mission. One such
mission, the Cosmic Origins Explorer (COrE)\footnote{\small
  http://www.core-mission.org}, has been proposed to ESA within Cosmic
Vision 2015-2025 \citep{2011arXiv1102.2181T}. Missions with similar
objectives, but different designs, have been proposed to NASA
\citep[see, e.g.,][]{2009AIPC.1141...10B}. Contamination by
foregrounds is one of the main worries for this measurement, and the
investigation of the severeness of the contamination, as well as the
development and validation of component separation methods adapted to
the challenge of measuring CMB $B$-modes, has focused significant
attention recently
\citep{2005MNRAS.360..935T,2006MNRAS.372..615S,2007PhRvD..75h3508A,
  2009A&A...503..691B,2009AIPC.1141..222D,2009MNRAS.397.1355E,2010MNRAS.408.2319S}.

Clearly, the WMAP mission lacks the sensitivity to place a strong
limit on the $B$-mode polarisation. While our result does not show any
detection of $B$-modes (see figure~\ref{fig:cmb-cl-bb} and
table~\ref{tab:binned_bbspec}), one of the points provided by the WMAP
collaboration\footnote{\small
  http://lambda.gsfc.nasa.gov/data/map/dr5/dcp/spectra/\\wmap\_bb\_spectrum\_9yr\_v5.txt},
in the first multipole bin, is significantly discrepant.

It is difficult for us to comment on the origin of this discrepancy,
as we do not exactly know how the WMAP error bars are obtained. It
may be that the WMAP data points published on the LAMBDA website
neglects noise correlations, or possible residual systematics or
foregrounds. The exact meaning of the posted error bars should
probably be clarified by the WMAP team. Our estimated error bar,
based solely on the dispersion of the data points used to generate the
binned power spectrum, seems to be more robust in that respect.

We are confident that our method, which has been also tested on
realistic simulations, is more effective for reducing foregrounds than
simple masking or template decorrelation. On simulations of future
observations with COrE, we have shown that it allows to reject
foreground contamination effectively enough to measure tensor to
scalar ratio of $10^{-3}$, limited by the sensitivity of the
observations rather than foreground contamination.
\section{Goodness of fit values} 
\label{sec:gof}
Finally, in order to demonstrate how well our measurements fit with
WMAP $9$-year best-fit $\Lambda CDM$ model, we have computed reduced
$\chi^2$ values to estimate the goodness of fit of our estimates per
multipole bin for each angular power spectrum. In principle a value of
goodness of fit per multipole bin equal to unity indicates that the
extent of the match between observations and estimates is in agreement
with the error variance. Table \ref{tab:gof} shows good compatibility
of our measured spectra with the best-fit cosmological model. The
goodness of fit values shown in the first two columns of this table
are obtained from the measured CMB power spectra and their errors for
all multipoles under consideration. However, the CMB likelihood is
non-Gaussian at large angular scales and
hence, the use of simple $\chi^2$ statistics is sub-optimal at such low
multipoles. Hence, the goodness of fit values, obtained from the same
measured power spectra and their errors for multipoles greater than
$23$, are tabulated in the last two columns of
table \ref{tab:gof} for comparison. In either case, our measurement
of $EE$ and $BB$ spectra is in significantly better agreement with the
WMAP best-fit model than that provided by WMAP collaboration, while
$TE$ and $TB$ are marginally discrepant with the WMAP best fit model,
with a reduced $\chi^2$ of order $1.4$ for $34$ degrees of freedom
(due in particular to a few points several sigmas away between $l=10$
and $40$). A complete investigation of this requires a more accurate
model of the measurement, and in particular of correlated errors, and
is beyond the scope of the present paper.
\begin{table}
  \caption{Comparison of goodness of fit values per multipole bin}  
  \centering \begin{tabular}{c c c c c c}
  \hline\hline 
  $XY$ & $NILC$ & $WMAP$ & $NILC$ & $WMAP$ \\ 
  $   $  & (all $l$)  &  (all $l$)  & ($l > 23$)  &  ($l > 23$)   \\ [1ex]\hline
  $EE$  & 0.86  &  2.21   & 0.87  &  1.81  \\ 
  $BB$  & 0.82  &  12.16  & 0.99  &  2.27  \\ 
  $TE$  & 1.48  &  0.94 & 1.58  &  0.83 \\ 
  $TB$  & 1.38  &  1.08  & 1.45  &  1.20  \\ 
  $EB$  & 0.95  &  $--$  & 0.99  &  $--$  \\ 
[1ex] \hline
\end{tabular} 
\label{tab:gof} 
\end{table}
\section{Conclusions}
\label{sec:conclusion}
In this work, we have computed CMB power spectra for polarised WMAP
observations. CMB polarisation maps are obtained from WMAP
observations using linear combinations that minimise the variance of
the recovered CMB on spherical wavelet (needlet) domains, that are
subsequently used to compute CMB power spectra.

Our analysis differs substantially from that of the WMAP team: we use
all WMAP channels, use a needlet ILC over the full range of harmonic
modes, and produce $9$ independent maps for each of $T$, $E$ and $B$,
from the different years of observation. Our estimated error bars do
not rely on a model of the WMAP noise, but instead are computed
directly from the estimated same-year power spectra ($9$ in our case)
and cross-year power spectra ($36$ in case of $EE$ and $BB$ spectra,
and $72$ in case $TE$, $TB$ and $EB$ spectra).

We find that our $EE$ power spectrum is in excellent agreement with
the expectations from the current cosmological model, while the
$9$-year WMAP $EE$ spectrum available publicly on the Lambda web
site\footnote{http://lambda.gsfc.nasa.gov} seems to be systematically
higher. Similarly, on very large scale, our $BB$ power spectrum is
consistent with zero, while the $9$-year WMAP $BB$ spectrum in first
multipole bin is not compatible with zero.

The agreement of our $TE$ and $TB$ measurements with the WMAP ones and
with the theoretical best fit model are good, but not perfect. The
origin of the discrepancy is not fully understood. Finally, our
measurements of the $EB$ spectra are compatible with zero, as expected
for a standard cosmological model.

\section*{Appendix}
\label{sec:appendix}
Suppose we have $M$ measurements (one measurement per year of
observation) of the full sky CMB, such that each of these measurements
contains the same signal which comprises the true CMB and residuals of
foreground emission and noise. The (residual) noise in these
measurements is statistically independent from year to year. The
harmonic coefficients of these measurements are expressed as,
\begin{eqnarray}
A^{X,I}_{lm}=S^{X}_{lm}+N^{X,I}_{lm} \hspace{0.2in} X=\{T,E,B\}
\hspace{0.2in}  I=1,..,M.
\end{eqnarray}
Here $A^{X}_{lm}$, $S^{X}_{lm}$ and $N^{X}_{lm}$ are the harmonic
coefficients of the measured CMB, of the sum of true CMB and residual
foreground, and of residual noise respectively. Since signal and residual
noise, and residual noise from year to year, are statically
independent, they obey the following relations,
\begin{align}
&E\left[S^{X}_{lm}N^{Y,I*}_{l^{\prime}m^{\prime}}\right]=0\\
&E\left[S^{X}_{lm}S^{Y*}_{l^{\prime}m^{\prime}}\right]=S^{XY}_{l}\delta_{ll^{\prime} }\delta_{mm^{\prime} }\\
&E\left[N^{X,I}_{lm}N^{Y,J*}_{l^{\prime} m^{\prime}}\right]=N^{XY}_{l}\delta_{ll^{\prime} }\delta_{mm^{\prime} }\delta_{IJ}
\end{align}
where, $S^{XY}_{l}$ and $N^{XY}_{l}$ are the angular power spectra of
signal and residual noise respectively.

From these $M$ maps $A^{X,I}_{lm}$, we have $M(M-1)$ cross-year
measurements $(\widehat{C}^{XY,IJ}, I \ne J)$ of the angular power
spectrum, such that each of them is an unbiased estimator of
$S^{XY}_{l}$ (although they are not independent). We have:
\begin{eqnarray}
\widehat{C}^{XY,IJ}_{l}=\frac{1}{2l+1}\sum^{l}_{m=-l}A^{X,I}_{lm}A^{Y,J*}_{lm}
\end{eqnarray}
\begin{eqnarray}
E\left[\widehat{C}^{XY,IJ}_{l}\right]=S^{XY}_{l} + N^{XY}_{l} \delta_{IJ}.
\end{eqnarray}
The average of all cross-year measurements of angular power
spectra is also an unbiased estimator of signal power spectrum
$S^{XY}_{l}$:
\begin{eqnarray}
\widehat{C}^{XY}_{l}=\frac{1}{M(M-1)}\sum^{M}_{I,J=1}\widehat{C}^{XY,IJ}_{l} (1-\delta_{IJ})
\end{eqnarray}
The variance of $\widehat{C}^{XY}_{l}$ is, by definition,
\begin{eqnarray}
V^{XY}_{l}=E\left[\left(\widehat{C}^{XY}_{l}\right)^{2}\right]-\left(E\left[\widehat{C}^{XY}_{l}\right]\right)^2
\label{equ:variance2}
\end{eqnarray}
where, $E\left[\widehat{C}^{XY}_{l}\right]$  and
$E\left[\left(\widehat{C}^{XY}_{l}\right)^{2}\right]$ are the
expectation values of $\widehat{C}^{XY}_{l}$ and
$\left(\widehat{C}^{XY}_{l}\right)^{2}$ respectively. We have:
\begin{align}
E\left[\widehat{C}^{XY}_{l}\right]=\frac{1}{M(M-1)}\sum^{M}_{I,J=1}E\left[\widehat{C}^{XY,IJ}_{l}\right](1-\delta_{IJ})=S^{XY}_{l}
\label{equ:est}
\end{align}
and
\begin{align}
E\left[\left(\widehat{C}^{XY}_{l}\right)^{2}\right]=\left[\frac{1}{M(M-1)}\right]^{2}\sum^{M}_{\substack{I,J,\nonumber\\K,L=1}}&E\left[\widehat{C}^{XY,IJ}_{l}\widehat{C}^{XY,KL}_{l}\right]\\
&\times (1-\delta_{IJ})(1-\delta_{KL})
\label{equ:est2}
\end{align}
In order to express
$E\left[\left(\widehat{C}^{XY}_{l}\right)^{2}\right]$ in terms of
$S^{XY}_{l}$ and $N^{XY}_{l}$, first we express the expectation value
of correlations among $(\widehat{C}^{XY,IJ})$ in terms of $S^{XY}_{l}$
and $N^{XY}_{l}$. We get:
\begin{align}
&E\left[\widehat{C}^{XY,IJ}_{l}\widehat{C}^{XY,KL}_{l}\right]=\left(S^{XY}_{l}\right)^2\nonumber\\
+&\frac{1}{2l+1}\left[\left\{\left(S^{XY}_{l}\right)^2+S^{XX}_{l}S^{YY}_{l}\right\}+\left(S^{XX}_{l}N^{YY}_{l}\delta_{JL}\right.\right.\nonumber\\
+&\left.\left. S^{YY}_{l}N^{XX}_{l}\delta_{IK}+S^{XY}_{l}N^{XY}_{l}\delta_{IJ}+S^{XY}_{l}N^{XY}_{l}\delta_{KL}\right)\right.\nonumber\\
+&\left.\left\{\left(N^{XY}_{l}\right)^2 \delta_{IL}\delta_{JK}+N^{XX}_{l}N^{YY}_{l}\delta_{IK}\delta_{JL}\right\}\right].
\label{equ:covar}
\end{align}
Then, we express $E\left[\left(\widehat{C}^{XY}_{l}\right)^{2}\right]$
in terms of $S^{XY}_{l}$ and $N^{XY}_{l}$ by combining equations
\ref{equ:est2} and \ref{equ:covar},
\begin{align}
E\left[\left(\widehat{C}^{XY}_{l}\right)^{2}\right] &\nonumber\\
=\left(S^{XY}_{l}\right)^2+&\frac{1}{2l+1}\left[\left\{\left(S^{XY}_{l}\right)^2+S^{XX}_{l}S^{YY}_{l}\right\}\right. \hspace{0.5in}\nonumber\\
+&\left.\frac{1}{M}\left(S^{XX}_{l}N^{YY}_{l}+ S^{YY}_{l}N^{XX}_{l}+2\,S^{XY}_{l}N^{XY}_{l}\right)\right.\nonumber\\
+&\left. \frac{1}{M(M-1)}\left\{\left(N^{XY}_{l}\right)^2+N^{XX}_{l}N^{YY}_{l}\right\}\right].
\label{equ:est2sn}
\end{align}
Finally, we express $V^{XY}_{l}$ in terms of signal and noise power
spectra, by combining equations \ref{equ:variance2}, \ref{equ:est} and
\ref{equ:est2sn}.
\begin{align}
V^{XY}_{l}=&\frac{1}{2l+1}\left[\left\{\left(S^{XY}_{l}\right)^2+S^{XX}_{l}S^{YY}_{l}\right\}\right. \hspace{1.0in}\nonumber\\
&\hspace{0.2in}+\left.\frac{1}{M}\left(S^{XX}_{l}N^{YY}_{l}+ S^{YY}_{l}N^{XX}_{l}+2\,S^{XY}_{l}N^{XY}_{l}\right)\right.\nonumber\\
&\hspace{0.2in}+\left. \frac{1}{M(M-1)}\left\{\left(N^{XY}_{l}\right)^2+N^{XX}_{l}N^{YY}_{l}\right\}\right].
\label{equ:variance}
\end{align}
In order to define an estimator for $V^{XY}_{l}$, first we define an
estimator $(\widehat{N}^{XY}_{l})$ for the residual noise power
spectrum $N^{XY}_{l}$.
\begin{align}
\widehat{N}^{XY}_{l}&=\frac{1}{M}\sum^{M}_{I=1}\widehat{C}^{XY,II}_{l}-\widehat{C}^{XY}_{l}\nonumber\\
&=\frac{1}{M}\sum^{M}_{I=1}\widehat{C}^{XY,II}_{l}-\frac{1}{M(M-1)}\sum^{M}_{I,J=1}\widehat{C}^{XY,IJ}_{l} (1-\delta_{IJ}).
\label{equ:auto_noise}
\end{align}
In equation \ref{equ:auto_noise}, the first term is an unbiased
estimator of the sum of the angular power spectra of signal and
residual noise, and the second term an unbiased estimator of the
angular power spectra of signal only.

Then, the estimator for $V^{XY}_{l}$ is constructed by replacing
$S_{l}$ and $N_{l}$ with the best-fit theoretical power spectrum
$C_{l,\rm th}$ and the estimated noise power spectrum
$\widehat{N}_{l}$ respectively in equation \ref{equ:variance}.
\begin{align}
  \widehat{V}^{XY}_{l}&=\frac{1}{2l+1}\left[\left\{\left(C^{XY}_{l,\rm
        th}\right)^2+C^{XX}_{l,\rm th}C^{YY}_{l,\rm th}\right\}\right. \nonumber\\
  &+\left.\frac{1}{M}\left(C^{XX}_{l,\rm th}\widehat{N}^{YY}_{l}+
      C^{YY}_{l,\rm th}\widehat{N}^{XX}_{l}+2\,C^{XY}_{l, \rm th}\widehat{N}^{XY}_{l}\right)\right.\nonumber\\
  &+\left. \frac{1}{M(M-1)}\left\{\left(\widehat{N}^{XY}_{l}\right)^2+\widehat{N}^{XX}_{l}\widehat{N}^{YY}_{l}\right\}\right].
\label{equ:est_variance}
\end{align}
In practice, the auto and cross angular power spectra are obtained
from observed NILC CMB maps after applying a mask. These angular power
spectra are corrected for the mask using the MASTER method
\citep{2002ApJ...567....2H} before using them to measure residual
noise power spectra. The estimated variance is divided by the sky
fraction $f_{\rm sky}$, as the effective number of modes for an
arbitrary multipole $l$, is now $(2l+1) f_{\rm sky}$ instead of
$(2l+1)$ (even if residual noise power spectra is corrected for the
mask).

$f_{\rm sky}$ is estimated as the average value of the product of the
two masks used, e.g. for $\widehat{C}^{TX}_{l}$ ($X$ being $E$ or $B$)
we use:
\begin{eqnarray}
f_{\rm sky} = \frac{1}{4\pi} \int m_T(\hat n) \, m_P(\hat n)\, d\Omega
,
\label{equ:fskyt}
\end{eqnarray}
where $m_T(\hat n)$ and $m_P(\hat n)$ are the masks for the
temperature and polarisation respectively (including noise weighting
modulation when necessary, i.e., for the noise-weighted estimates the
masks are real-valued, not just binary masks). For
$\widehat{C}^{XY}_{l}$ ($X$ and $Y$ being each either $E$ or $B$) we
use:
\begin{eqnarray}
f_{\rm sky} = \frac{1}{4\pi} \int \left[ m_P(\hat n) \right ]^2 \,
d\Omega.
\label{equ:fskyp}
\end{eqnarray}

\section*{Acknowledgements}
Soumen Basak is supported by a `Physique des deux infinis' (P2I)
postdoctoral fellowship. We acknowledge the use of the Legacy Archive
for Microwave Background Data Analysis (LAMBDA). Support for LAMBDA is
provided by the NASA Office of Space Science. The results in this
paper have been derived using the HEALPix package
\citep{2005ApJ...622..759G}. The authors acknowledge the use of the
Planck Sky Model \citep[PSM,][]{2012arXiv1207.3675D}, developed by the
Planck working group on component separation, for making the
simulations used in this work. We thank Jean-Fran\c{c}ois Cardoso,
Guillaume Castex, Eiichiro Komatsu, Maude Le Jeune, Mathieu
Remazeilles and Radek Stompor for useful discussions. We also wish to
thank the anonymous referee for useful comments that helped improve
our analysis and manuscript.

\label{lastpage}

\end{document}